\documentclass[journal,comsoc]{IEEEtran}

\usepackage[T1]{fontenc}
\usepackage{cite}
\usepackage{amsmath,amssymb,amsfonts}
\usepackage{algorithmic}
\usepackage{graphicx}
\usepackage{textcomp}
\usepackage[dvipsnames]{xcolor}

\usepackage{subfigure}
\usepackage{enumitem}
\usepackage{array}
\usepackage{multirow}
\usepackage{footnote}
\usepackage{color, colortbl}
\usepackage[hidelinks]{hyperref}

\newcolumntype{M}[1]{>{\centering\arraybackslash}m{#1}}
\newcolumntype{C}[1]{>{\centering\arraybackslash}p{#1}}

\newcommand\hl[1]{%
  \bgroup
  \hskip0pt\color{red!80!black}%
  #1%
  \egroup
}

\newcommand\cm[1]{%
  \bgroup
  \hskip0pt\color{blue!80!black}%
  #1%
  \egroup
}

\def\BibTeX{{\rm B\kern-.05em{\sc i\kern-.025em b}\kern-.08em
    T\kern-.1667em\lower.7ex\hbox{E}\kern-.125emX}}
    
\hypersetup{
	pdftitle={Characterization and Identification of 5G Performance Bottlenecks: An Experimental Study},
	pdfauthor={Georgios Patounas, Xenofon Foukas, Ahmed Elmokashfi, Mahesh K. Marina},
	pdfsubject={Characterization and Identification of 5G Performance Bottlenecks},
	bookmarksnumbered=true,     
	bookmarksopen=true,         
	bookmarksopenlevel=1,       
	pdfpagemode=UseOutlines
}
    
\begin{document}

\title{Characterization and Identification of Cloudified Mobile Network Performance Bottlenecks}

\author{Georgios~Patounas\IEEEauthorrefmark{1},
 Xenofon~Foukas\IEEEauthorrefmark{2},
 Ahmed~Elmokashfi\IEEEauthorrefmark{1},
 Mahesh~K.~Marina\IEEEauthorrefmark{3}
 \thanks{\IEEEauthorrefmark{1} G. Patounas and A. Elmokashfi are with the Simula Metropolitan CDE, Oslo, Norway (e-mail: \{gpatounas, ahmed\}@simula.no).}
 \thanks{\IEEEauthorrefmark{2} X. Foukas was with The University of Edinburgh for the duration of this work. He is now with Microsoft Research, Cambridge, United Kingdom (e-mail: xefouk@microsoft.com).}
 \thanks{\IEEEauthorrefmark{3} M. K. Marina is with The University of Edinburgh (e-mail: mahesh@ed.ac.uk).}}

\maketitle

\begin{abstract}
This study is a first attempt to experimentally explore the range of performance bottlenecks that 5G mobile networks can experience. 
To this end, we leverage a wide range of measurements obtained with a prototype testbed that captures the key aspects of a cloudified mobile network. We investigate the relevance of the metrics and a number of approaches to accurately and efficiently identify bottlenecks across the different locations of the network and layers of the system architecture. Our findings validate the complexity of this task in the multi-layered architecture and highlight the need for novel monitoring approaches that intelligently fuse metrics across network layers and functions. In particular, we find that distributed analytics performs reasonably well both in terms of bottleneck identification accuracy and incurred computational and communication overhead.   
\end{abstract}


\section{Introduction}
\label{sec:intro}

\IEEEPARstart{T}{he} journey towards 5G has been driving innovation in mobile networks in the recent past. 
Compared to previous generations of mobile networks, the most striking and perhaps the defining aspect of 5G from an architectural perspective is the underlying vision to turn it into a multi-service infrastructure. 
In other words, to not be limited to the mobile broadband and voice service (the {\em raison d'\^etre} of mobile networks till date) and instead become the basis for a wide range of services including various Internet of Things (IoT) applications and mission critical services like public safety communication.
To cost-effectively realize this vision, the concept of network slicing has emerged as the way forward~\cite{5gppp}. 
Essentially the idea is to virtualize the physical infrastructure underlying the mobile network, and realize each service as a suitably customized end-to-end composition of virtual network functions (VNFs) making up an isolated logical instance, called a slice, over the shared virtualized infrastructure. 

As this would require embracing some of the key technologies from the cloud computing domain such as infrastructure virtualization, Network Functions Virtualization (NFV) and Software-Defined Networking (SDN), 5G can be seen as ``cloudification'' of mobile networks.
While this cloudification is considered as they main hallmark of 5G, several operators have already embraced it in their 4G networks.

Service assurance is key to the success of the 5G vision. Without service quality guarantees, it would be hard to incentivize emerging services (e.g., connected vehicles) and existing services with their own custom network infrastructure (e.g., public safety) to come under the 5G fold. 
Being able to provide such guarantees however, hinges on firstly understanding the factors (i.e., potential service performance bottlenecks) that can influence service quality. 
Such understanding can help shape the design of suitable monitoring systems to efficiently track the quality experienced by services over the shared infrastructure and in turn guide dynamic control of resource allocations to ensure each service meets its requirements. 

{\em Our focus is on characterizing and identifying service performance bottlenecks in mobile networks with an emphasis on next generation architectural paradigms, which to the best of our knowledge has not been addressed to date.} 
Doing so, however, is not straightforward given that a cloudified infrastructure is inherently multi-layered, comprising of infrastructure, network function and service layers along with a cross-cutting management \& orchestration (MANO) layer.
Performance bottlenecks can span and be impacted by all these layers and to make matters more complicated, multiple bottlenecks could manifest with seemingly identical effects on the end-to-end service quality that the users experience as well as on the aggregated telemetry that is available to the network operator.
This difficulty in analyzing and attributing bottlenecks spanning several physical and logical layers contrasts cloudified mobile networks with traditional 3G and 4G networks.
While the architectures at a conceptual level are becoming less complex, a single network function is stretched across several layers spanning all the way from the bare metal to the service level.

The individual domains that make up this complex ecosystem have been previously studied: mobile networking; data centers and cloud computing; virtualization of functions and performance monitoring; anomaly detection and root cause analysis systems~\cite{calero2015monpaas,moshref2016trumpet,baranasuriya2015qprobe,sciancalepore2018z}. However, the majority of these studies are mostly confined to their particular domain of focus. 
In a cloudified, mobile network setting which brings together all these different dimensions, the challenges of pinpointing performance bottlenecks while taking into account all involved domains, is a previously uncharted territory.

In view of the above, in this paper we present an experimental study on characterizing service performance bottlenecks as relevant to cloudified mobile networks. 
This study is enabled by a prototype testbed we deployed.
Utilizing this testbed, we study the impact of bottlenecks at various points along the service path and across the different layers of the system architecture. 
We evaluate the results with end-to-end service quality in mind. 
Given that the number and type of system parameters 
measured have direct implications on the efficiency and efficacy of monitoring systems, we examine the value of various monitoring parameters spanning the whole system and approaches to monitoring in terms of their ability to efficiently and accurately pinpoint different bottlenecks.

Our study allows us to make the following key observations:
\noindent 1) Different types of bottlenecks originating from different parts and layers of the system architecture can have seemingly similar effects in terms of the observed performance, making the identification of bottlenecks a challenging task.

\noindent 2) While techniques like manual inspection of Key Performance Indicators (KPIs) can provide insights and narrow down the potential causes affecting performance, the number of sources and volume of monitoring data as well as the complexity of their mutual correlations make this approach impractical if not impossible for most realistic settings.

\noindent 3) Machine learning (ML) algorithms can greatly automate the process of analyzing measurements. Leveraging a diverse set of features extracted from measurements at multiple layers of the emerging cloudified mobile network system is extremely important and makes it possible to distinguish different bottlenecks as well as different profiles of similar bottlenecks that can manifest in greatly distinct ways. 

\noindent 4)There is a trade-off between the granularity at which bottlenecks are described and the ability to accurately and efficiently identify them.
Interestingly, identification based on coarse bottleneck types can offer reasonable accuracy provided a comprehensive set of measurements is available. Finer granularity enhances the accuracy further and increases troubleshooting efficiency by narrowing down the root causes. However, granularity comes at the cost of overhead for the definition of bottlenecks which can become intractable in complex networks.

\noindent 5) Complex networks will produce complex bottlenecks. Identification of bottlenecks composed of different causes and presented in different locations is a difficult task. Distributed analytics can yield excellent accuracy, even for composite bottlenecks while simultaneously keeping the computational and communicational overhead, an ever-important concern, low. 

\bigskip
We should clarify that our intention with this work is not to exhaustively enumerate and examine all potential performance bottlenecks, but rather to empirically highlight key challenges that arise when instrumenting and monitoring a cloudified mobile network architecture for dynamic root cause analysis. In doing so, we map the problem space and uncover promising ideas and solutions that will inform the design of effective monitoring systems for mobile networks of today and beyond. This is a noteworthy and unique aspect of our work. We believe that this would prove valuable, as a concrete empirical case study, to research efforts going forward.

The rest of the paper is structured as follows. 
Section~\ref{sec:background} provides the relevant and essential background concerning components and elements of cloudified mobile networks. Section~\ref{sec:methodology} describes our experimental methodology and testbed setup. 
Section~\ref{sec:analysis} examines different data-driven approaches to discriminating among service performance bottlenecks. 
Section~\ref{sec:bottlenecks} approaches the problem from a practical network monitoring perspective.  
Section~\ref{sec:composites} examines the complexities of multiple simultaneous bottlenecks.
Section~\ref{sec:discussion} discusses issues of measurements selection and overhead minimization as well as the lessons learned.
Section \ref{sec:related} discusses related work.
Finally, conclusions are provided in Section \ref{sec:conclusions}.
\section{background}
\label{sec:background}

\begin{figure}[t]
	\centering	
	\includegraphics[width=1\columnwidth]{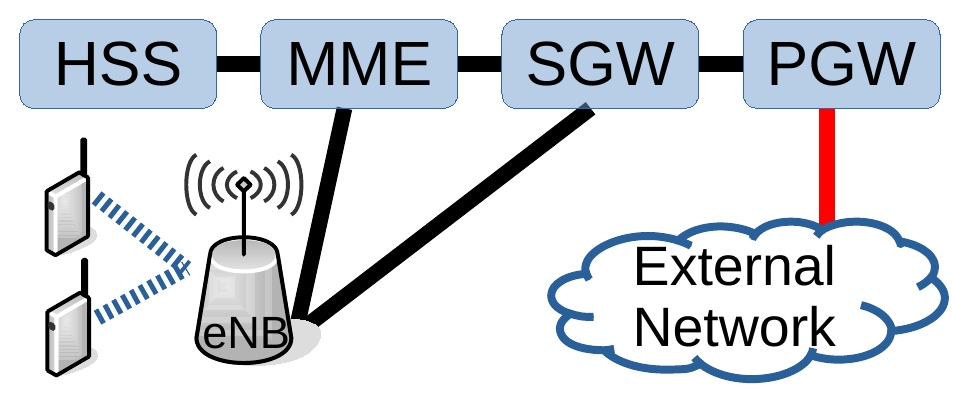}
	\caption{A generic representation of the 4G network architecture.}
    \label{fig:4g_architecture}
    \vspace{-4mm}
\end{figure}

Generally speaking, mobile networks are composed of two main components: the Core Network (CN) and the Radio Access Network (RAN).
Considering currently deployed 4G networks as an example (see Fig.~\ref{fig:4g_architecture}, the CN, referred to as evolved packet core (EPC) in 4G, contains several entities including: the Packet Data Network Gateway (PGW) which acts as the gateway between the mobile and external networks; the Serving Gateway (SGW) which manages user equipment (UE) context and acts as its mobility anchor; the Mobility Management Entity (MME) which performs control actions such as SGW selection, UE paging and tagging, etc.; and the Home Subscriber Server (HSS) which manages registration of users on the network and functionalities like policy enforcement and charging.
The CN has traditionally been realized via dedicated, high-performance hardware that is distributed in a small number of locations that cover wide areas with millions of users (e.g., 10-20 per provider in the US). 
On the other hand, the RAN is tasked with providing the wireless access to users and a host of other procedures needed to ensure that each user can join the network, maintain connectivity over time and use the network services. The RAN consists of thousands of base stations that carry the necessary wireless, compute and networking equipment in a unit called eNodeB (eNB) in the 4G architecture. As with the CN, the RAN and eNBs are also traditionally built on specialized hardware geographically distributed to provide the necessary coverage and performance to the users throughout the network (i.e., high density is needed in urban areas).

The fact that the network infrastructure consists of specialized hardware, means that it needs to be statically provisioned and is not easily modifiable. 
The inherent inflexibility of this approach along with the spatio-temporal volatility of users and their traffic mean that the task of provisioning resources efficiently and sufficiently is extremely difficult.
In addition, to support the ever increasing performance and functionality requirements, the architecture has become complex and the deployed infrastructure immense. 
The strain placed on the network is expected to grow almost exponentially in the coming years as providers are constantly working to capture new markets with the next generation of 5G mobile networks.
Three key use case families have been identified for 5G: enhanced mobile broadband; massive machine type communications and critical communications.
Several applications can be served within those ranging from generic broadband access to specialized networks for sensors (e.g., smart meters) or those requiring low latency or high reliability (e.g., connected vehicles, factories). 
The vastly different requirements of these applications cannot be satisfied by a one-size-fits-all architecture giving rise to technological and architectural proposals to accommodate them~\cite{5gppp, tr_ngmn_5gwhitepaper, itu_imt2020, foukas2017mcom}.

\begin{figure}[t]
	\centering	
	\includegraphics[width=1\columnwidth]{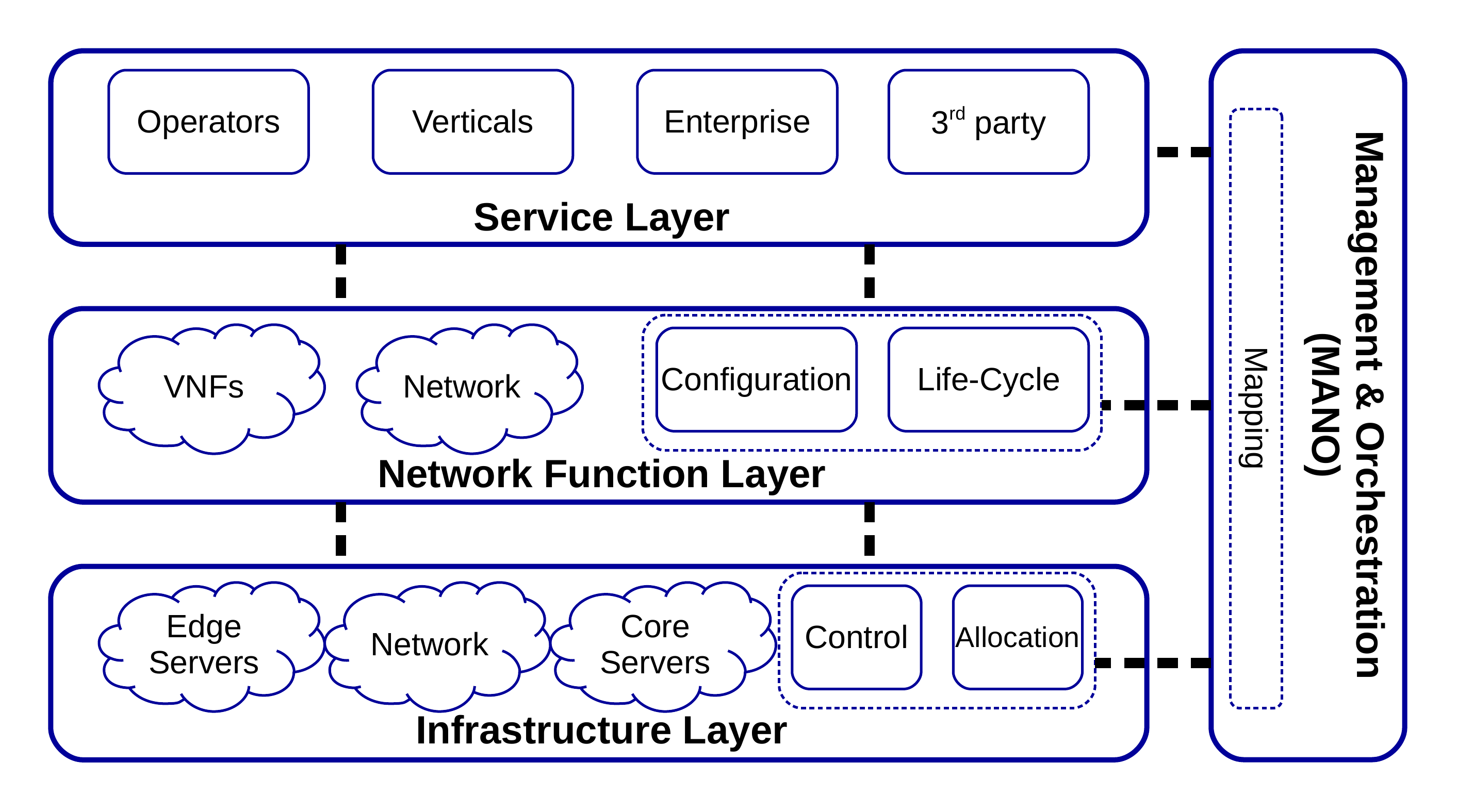}
	\vspace{-10mm}
	\caption{Framework representing the leading 5G architectural proposals.}
    \label{fig:architecture}
    \vspace{-4mm}
\end{figure}

The problems of inflexibility and complexity of the network can conceptually be addressed by adopting technologies that have long been used in the cloud computing context.
The need for native support of softwarization (i.e. moving from rigid physical network function (PNFs) to virtual network functions (VNFs) and leverage SDN for consolidated control separated from data plane) and network slicing or multi-tenancy have been widely established, with the main 5G architectural proposals built around this as shown by the generic framework presented in Fig.~\ref{fig:architecture}.
The framework is composed of three main layers: the infrastructure layer, the network function layer and the service layer. 
The infrastructure layer broadly refers to the physical network infrastructure spanning both the RAN and the CN  i.e., network, compute, storage and memory. The network function layer consists of the actual network functions (VNFs and PNFs) that are composed to realize a service (or a network slice). The service layer comprises all elements needed for linking actual businesses (business models) with network slices. A newly introduced Management and Orchestration (MANO) entity is tasked with the overall life-cycle management of network slices that includes: creation of network slices as needed via mapping across the different layers between service specifications, NFs and infrastructure resources; their (re-)configuration assisted by monitoring throughout the lifetime of each service.

\begin{figure}[t]
	\centering	
	\includegraphics[width=1\columnwidth]{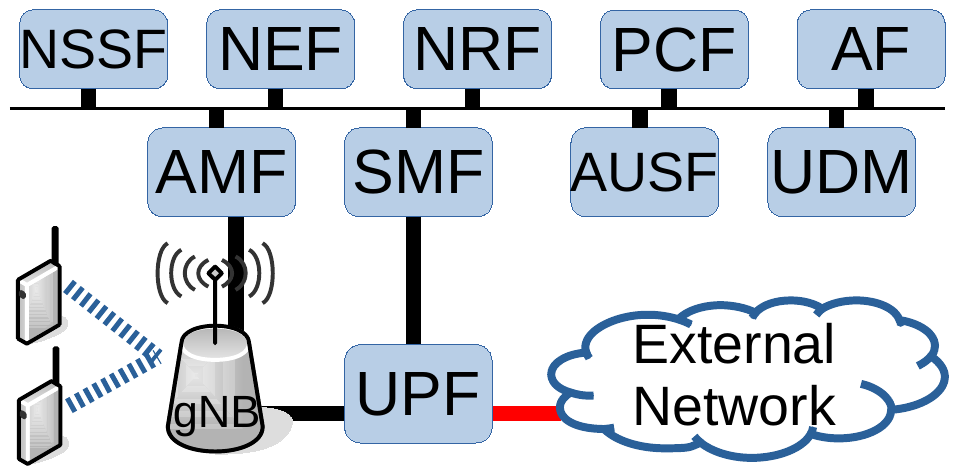}
	\caption{A generic representation of the 5G network architecture.}
    \label{fig:5g_architecture}
    \vspace{-4mm}
\end{figure}

Further to the described framework, the 5G specifications bring marked changes to the spectrum, signal processing, communication protocols and functionality of the network. However, one can clearly identify the functionality provided by the NFs of the 4G network, in their evolved form as VNFs (see Fig.~\ref{fig:5g_architecture}). The gNB takes the place of the eNB. The Authentication Server Function (AUSF) and User Data Management (UDM) fulfill the duties of the HSS. The Access and Mobility management Function (AMF) takes the place of the MME. Finally, the Session Management Function (SMF) and User Plane Function (UPF) take over the duties of the SGW and PGW while splitting the user plane and the control plane, in the interest of disaggregation and flexibility. For the same reasons, a number of new VNFs are defined as well that can vary depending on the exact network configuration (omitted for brevity).

\section{Experimental Methodology}
\label{sec:methodology}

This section presents our experimental methodology including the testbed, the performance bottlenecks considered and the methodology to produce them. 

\subsection{Testbed Setup}
\label{sec:methodology_testbed}
\subsubsection{Network functions considered.} 
In this paper, we focus on the performance implications of bottlenecks from the viewpoint of a single network slice. Accordingly, we deploy and chain all the VNFs required to create an end-to-end network slice. For the slice creation on the RAN side, we leverage the Orion RAN slicing system~\cite{foukas2017orion}.   
Orion provides functionally isolated virtual control planes for network slices and reveals virtualized radio resources to them through a Hypervisor component, ensuring both functional and performance isolation.
Based on Orion's design, each tenant can take full control of its slice by being assigned its own RAN controller, which can be fully configured in terms of the control operations from the MAC layer and above. 
Orion is in turn built on top of the open source OpenAirInterface (OAI) 4G LTE platform~\cite{nikaein2014openairinterface}, which provides the implementation of the eNB. 
For the CN, we employ openair-cn~\cite{openair_cn}, which is one of the most complete open source EPC implementations available, allowing the deployment of the HSS, MME and SPGW functions as separate processes over different physical or Virtual Machines (VMs).
The VNFs were implemented as full VMs however container based virtualization with the use of Linux containers or Docker is possible with minimal modifications and no impact on the study.

\subsubsection{Physical testbed configuration.}
\begin{figure}[t]
	\centering
	\includegraphics[width=1\columnwidth]{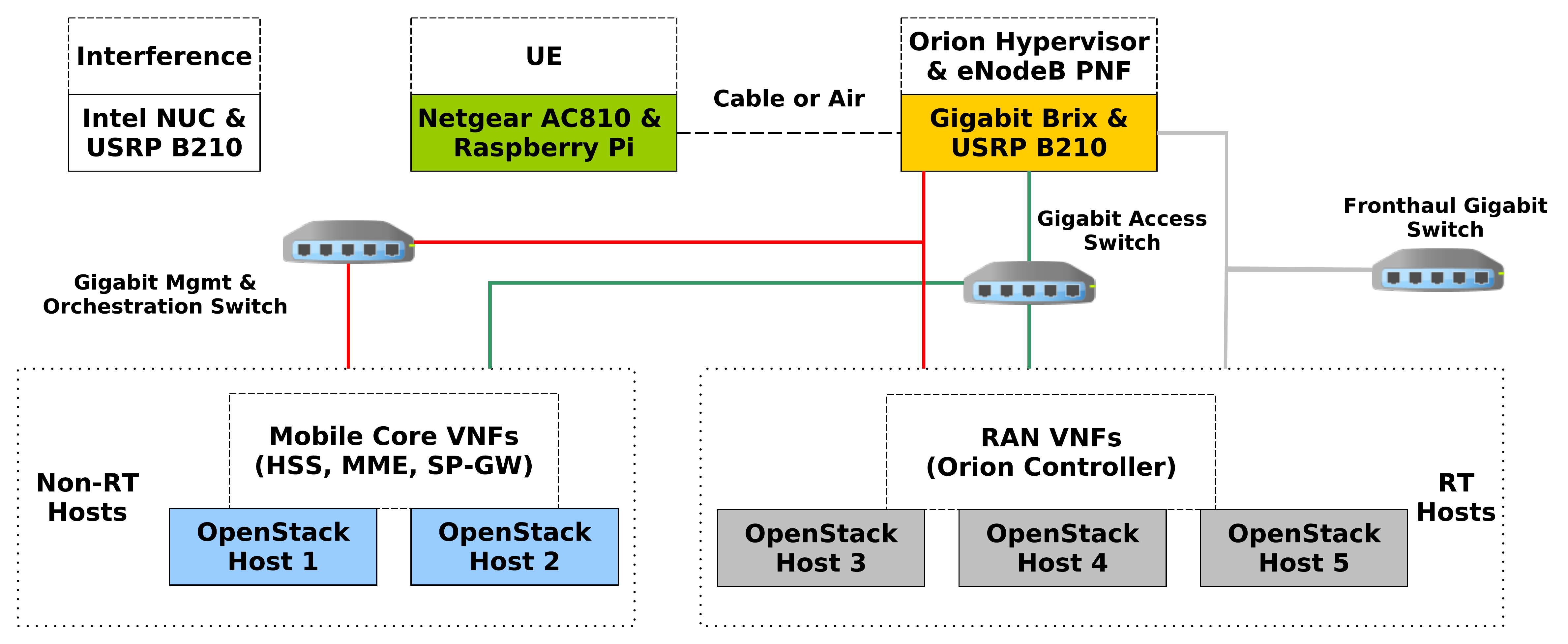}
	\vspace{-5mm}
	\caption{Schematic of the testbed setup.}
	\label{fig:openstack_testbed}
\end{figure}

We setup an OpenStack-based prototype testbed for our experimentation, which as illustrated in Fig.~\ref{fig:openstack_testbed}, provides 5 compute nodes. 
Two of the nodes (24-core Intel Xeon Silver, 64GB RAM) were used for the deployment of the mobile core VNFs (HSS, MME, SPGW), which do not present strict execution or latency requirements. Three of the nodes (10-core Intel Xeon Silver, 16GB RAM) were used for the deployment of RAN-related VNFs with real-time constraints (e.g., the RAN slice controller). 
A number of optimizations were applied to these hosts to enable real-time performance (disabled CPU C-states/frequency-scaling, low-latency Linux kernel, isolated CPUs for the operation of the hosts' OS and CPU pinning for the deployed VNFs).

For the Orion Hypervisor and eNB, we used a Gigabyte Brix Pro (4-core Intel i7, 8GB RAM) connected to a USRP B210 Software-Defined Radio (SDR) unit, acting as a PNF. 
Another PC (Intel NUC), with similar characteristics to the Brix Pro, and a B210 SDR were used for the experiments with an interference source. Finally, for the user equipment, we used a Netgear AC810-100EUS with a Raspberry Pi 3 Model B+ attached to it via USB and configured in IP passthrough mode. The Netgear unit attaches to the mobile network and the Raspberry Pi is used as a terminal (traffic endpoint). To ensure the fidelity and reproducibility of our experiments, we used a cabled connection between the Netgear unit to the eNodeB SDR, apart from interference-related experiments that require an over-the-air connection.

This testbed, although small in scale, captures all key aspects of a cloudified architecture including 5G RAN slicing and the use of VNFs (see \cite{foukas2018testbed} for further details on the testbed). 

\subsection{Testbed scope}

The design of the testbed is largely influenced by the paradigms exemplified by the 5G architecture as described in Sec.~\ref{sec:background}.
Nevertheless, the trend of cloudification of networks has existed for some time with selected NFs being virtualized and recently deployments of end-to-end virtualized 4G networks~\cite{4grakuten, 4gtmobile}.
In much the same way, we implement 4G components as VNFs over datacenter level infrastructure. Additionally we utilize software that enables network slicing, introducing the function of an eNB controller.
Hence, our testbed faithfully captures a cloudified mobile network with elements of network slicing.
While this foregoes the disaggregation of functions possible with the 5G specifications and 5G radio \cite{3gpp_release15}, it allows us to evaluate bottlenecks on a testbed that represents the challenges of a mobile network and the additional complexity of virtualized infrastructure and a multi-layered architecture.

\subsection{Bottlenecks Considered}
\label{sec:methodology_bottlenecks}

We consider five categories of performance degradation that commonly plague wireless communication, networks and virtualized infrastructure, with the aim of uncovering the range of bottlenecks that could affect end-to-end service quality of a mobile network. While these are performance issues that can affect most networks, we highlight the challenges created by the evolution to cloudified mobile architectures.

\begin{description}[style=unboxed,leftmargin=0cm]
\item[Interference.] Besides well established issues of existing cell-edge interference~\cite{boudreau2009interference} mobile networks are evolving in ways that will provide new and dynamic sources of interference. 
The sheer amount of devices, along with the different requirements created by network slicing and IoT technologies bring about new demands for the wireless base-stations. At the same time, the introduction of small-cells with unpredictable operational parameters presents additional challenges in operations ~\cite{lopez2011enhanced}.
To emulate such a bottleneck, we use an SDR unit, which, through GNURadio~\cite{gnuradio}, creates interference to match the downlink frequency and bandwidth of the eNB.

\item[Packet Loss.] Packet loss can be attributed to a number of different causes such as link congestion, software or hardware faults and misconfigurations which in turn can induce bit errors, buffer overruns or intentional drops in any node of the network~\cite{netpilot}. 
With the introduction of ultra-reliable services~\cite{5gppp}, loss of packets becomes a critical bottleneck.
To emulate loss, we use the netem tool~\cite{netem} and experiment with varying degrees of packet loss in different parts of the service (slice) path.

\item[Link Congestion.] As the mobile architecture is moving towards shared infrastructure and convergence with the Internet is expanding, link utilization may become unpredictable~\cite{akyildiz2015softair}. 
Especially in deployments where several slices are sharing the same links to serve their customers and traffic may be routed through public networks or commodity datacenters, management of link capacity can be challenging.
To emulate congestion we use the iperf3 tool~\cite{iperf3}, which allows us to generate traffic in one or both directions of various links within the network during our experiments.

\item[Computational Resources.] Another effect of shared infrastructure is that VNFs may end up competing for shared computational resources. 
Despite the significant research on performance isolation in virtualized environments~\cite{bachiega2018container}, it is far from a resolved problem with solutions usually offering a compromise between isolation, resource optimization and elasticity (e.g.~\cite{zhao2017quantifying}).  
In our setup, competition for the computational resources is achieved by use of the stress tool~\cite{stress}, which comes with a number of functions to strain the CPU, memory and storage.

\item[Delay.] Arguably one of the most ambitious targets set for next generation networks, is the support for low-latency services~\cite{5gppp}. While more efficient efficient PHY and MAC techniques aim to reduce the delay of the network, the need for services delivered to the users with extremely low end-to-end latency also imposes strict requirements on VNF placement and performance, making the ever important end-to-end delay especially critical.
Using netem and an approach similar to the packet loss experiments, we emulate a wide range of delay related scenarios.

\end{description}

\medskip
It is important to note, that while the above bottlenecks can manifest independently, there are inter-dependencies that will often lead to cascading effects depending on their severity.
Interference can trigger packet loss and link congestion by affecting the available throughput of the wireless link. 
Inadequate computational resources can cause increased delay and in extreme cases, packet loss.
Congested links can trigger packet loss and increase delay.
While the inter-dependencies are important, in the course of our experiments we examined each bottleneck separately to capture their core behaviour.
Note that this isolation is desirable, since being able to identify and separate scoped causes of performance degradation is a reasonable starting points towards localizing and attributing those with complex behaviour.   
Although these broad bottlenecks exist in traditional mobile networks as well as other types like fixed or data center networks, it is clear that a cloudified mobile network brings new dimensions to them.  
This makes the scenarios we are testing unique in that we are examining bottleneck impact on: 1) end-to-end performance, which includes the RAN thus making it specific to mobile networks,  2) a fully softwarized slice, thus making it relevant to cloudified mobile networks.

\subsection{Bottleneck Profiles}

In this work, we consider a number of different bottleneck profiles, summarized in Table~\ref{table:2}, that cover the five categories of performance degradation above.
A profile is characterized by the severity and location of the bottleneck e.g. 1\% packet loss at the SPGW.
Each profile is the sum of five independent runs, each being two-minutes long i.e. the bottleneck was induced in multiple runs to ensure no transient conditions unrelated to the examined bottleneck affected the measurements.
In Sections~\ref{sec:analysis}~and~\ref{sec:bottlenecks} up to five profiles for each bottleneck category are examined, amounting to a total of seventeen distinct profiles. 
The profiles were selected by considering realistic bottleneck scenarios that cover the important nodes and links of the system and are as follows.
Profiles of interference were emulated by producing downlink interference at two distinct power output levels.
Packet loss is produced in all major functions: the controller, the eNB and the SPGW at three different intensities. 
Congestion was produced on the link between the eNB and controller as well as the one between the external network and SPGW.
Computational resources were stressed on the SPGW VNF, its host and the controller.
Delay was introduced on the SPGW where large variability of delay can be tolerated, as well as the controller and the eNB where an extremely low link latency is required.
Finally, in addition to these singular bottlenecks, in Section~\ref{sec:composites} we introduce three composite bottleneck profiles  (see Table~\ref{table:3}). 

\subsection{Data Collection and Metrics}
\label{sec:methodology_metrics}

Our measurements are organized in three groups, corresponding to the layers of the 5G mobile network as outlined in Fig.~\ref{fig:architecture}, namely the service layer, the network function layer and the infrastructure layer.
Measurements are logged at all the VNFs and PNFs of the mobile network as well as the UE that was the sink of user traffic during our experiments and the source server located outside the mobile network. 
We list the measurements and tools used below.

\begin{description}[style=unboxed,leftmargin=0cm]
\item[Service layer.] The D-ITG tool~\cite{ditg} is used to create the data-plane traffic in the mobile network. 
TCP streams are created on the downlink, with the server (in the testbed but external to the mobile network) sending data to the UE. 
Using, D-ITG, the throughput and Round-Trip Time (RTT) of the traffic streams are continuously logged at the two endpoints.

\item[Network Function layer.] Custom logging is used in the controller to utilize the Orion controller's statistics manager. 
This logs statistics for Radio transmissions (TX) and retransmissions (RTX) which identify the exact number of MAC frames transmitted and re-transmitted over the wireless link, respectively. 
The current Channel Quality Indicator (CQI) and Modulation and Coding Scheme (MCS) are also logged in the controller. CQI is a measure of wireless channel quality incorporating, among others, signal to noise and interference ratio.
MCS is a physical layer parameter selected by the eNB in close relation to the CQI and the amount of data available for transmission. 
We note that our setup only involves Single-Input Single-Output configurations and thus we do not consider additional measurements like rank indicators and PMI of UEs i.e. measurements pertinent to the multiple-input, multiple-output configuration.

Further, custom logging in the eNB tracks scheduling decisions and missed scheduling deadlines (MSD), which capture the control functions of the controller and eNB that decide allocation of resources to UEs.   
Finally, the SAR tool~\cite{sar} continuously monitors TCP retransmissions on all VNFs and the UE.

\item[Infrastructure layer.] The SAR tool is used to continuously monitor activity counters of the operating system (CPU, memory, storage, network TX and RX) in the physical hosts, the VNFs (SPGW, HSS, MME, eNB, Orion controller) and the UE. In addition, delay measurements are obtained for three key links of the system: The controller link between the eNB and the Orion controller, the S1-U link between the eNB and SPGW, the SGi link between the SPGW and the external network.
\end{description}

\medskip
The complete number of monitored KPIs is 52 and can be categorized by layer as seen in Table~\ref{table:1}. Six (TCP retransmissions, CPU, memory, storage, network TX and RX) are taken at each of seven locations (the HSS, MME, SPGW, SPGW host, eNB, controller and UE) providing forty-two measurements. Three measurements for link delay are provided by the three monitored links. Seven custom measurements (CQI, MCS, MSD, Radio TX and RX, throughput, RTT) are taken at appropriate locations (controller, eNB, UE).
\section{Are different bottlenecks easily separable?}
\label{sec:analysis}

Today, network operators leverage an array of approaches to identify root causes of performance degradation.
These include the visual inspection of measurements, correlation among various metrics and more recently machine-learning based systems that attempt to classify or partition relevant measurements~\cite{iyer2017automating, PadmanabhaIyerTradeoff}.
In this section, we explore whether common troubleshooting approaches are able to distinguish between various bottleneck profiles.

\subsection{Visual Inspection}
\label{sec:analysis_initial}

We begin by visually inspecting each measurement over the range of our experiments to determine their utility in identifying the bottlenecks.
In the course of visual inspection, a measurement is potentially useful if it provides a marked variation compared to its baseline value when encountering a specific bottleneck profile.
However, this potential usefulness can quickly diminish if the same measurement provides similar variations for a range of different profiles.
Considering the above, we analyze a subset of the profiles, to exemplify two challenges of discriminating among bottlenecks: the same bottleneck may occur at different locations in the network producing significantly different ``signatures'' (i.e. the observed effect to the service quality) and conversely, different bottlenecks manifesting at the same location can produce similar signatures hindering the identification of the underlying problem.
We should stress here that the observed complexities of bottleneck identification are by no means an isolated phenomenon and are in fact the norm when considering bottlenecks in complex networks.
We have visually inspected all of the profiles to find the described challenges to be very common. 

Here, we examine the following: 
(profile 02) severe interference on the wireless link between the eNB and UE; (profile 03) packet loss at the SPGW; (profile 07) packet loss at the eNB; (profile 08) congestion on the link between the SPGW and the external server (in 3GPP terminology the SGi interface) -- see Table~\ref{table:2} for further details.
In particular, we examine whether these profiles are visually distinguishable, from the baseline of our testbed as well as from each other based on the layered measurements outlined in Sec.~\ref{sec:methodology_metrics}.

\begin{figure}[t]
	\centering
	\includegraphics[width=1\columnwidth]{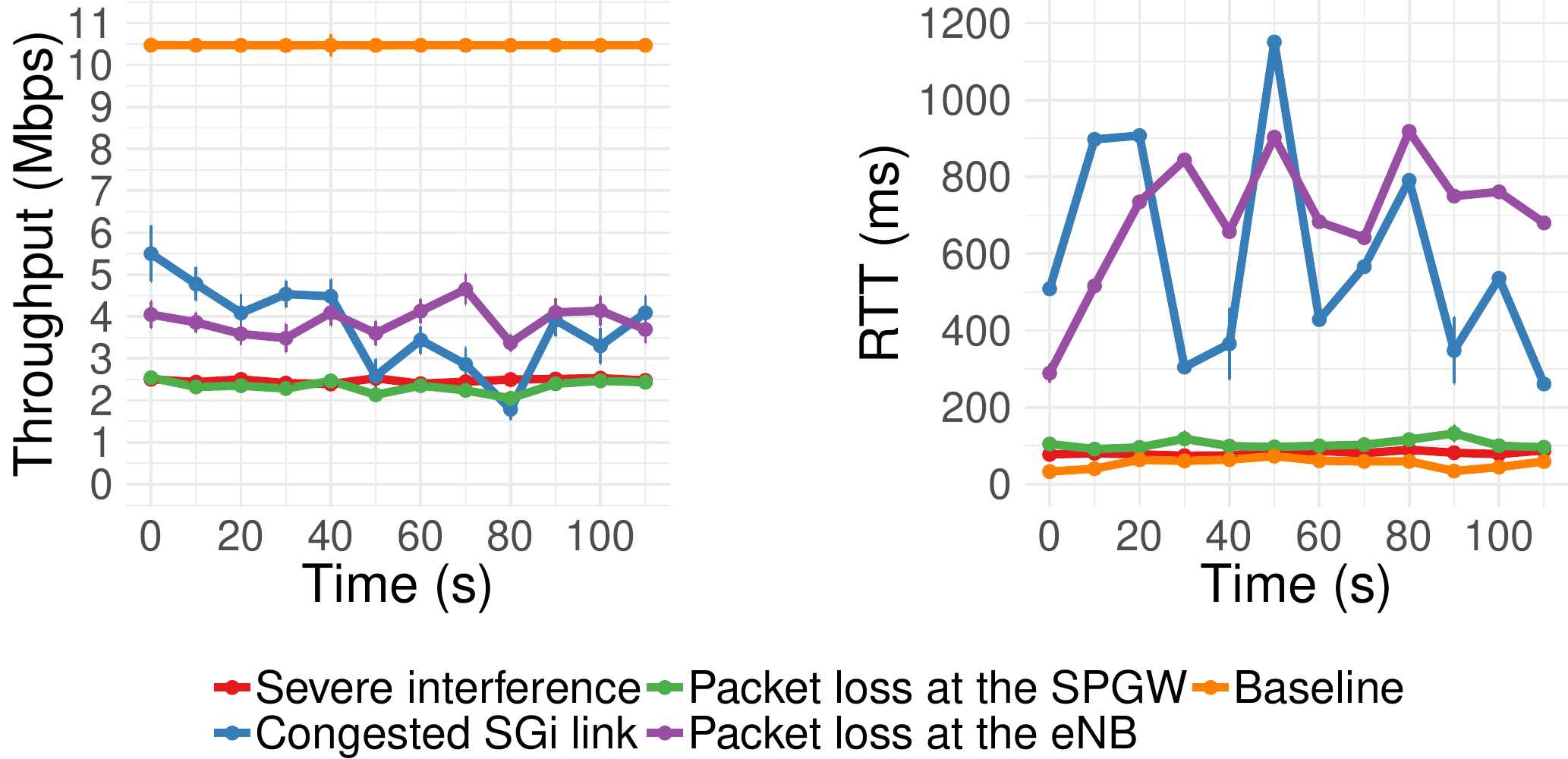}
	\vspace{-8mm}
	\caption{Service layer measurements: End-to-end throughput and round-trip time (RTT).}
	\label{fig:third_set_service}
	\vspace{-4mm}
\end{figure}

\noindent{\bf Service layer.} Fig.~\ref{fig:third_set_service} shows the throughput and RTT experienced by the UE for the selected four profiles and the baseline. 
Here, it is evident that different bottlenecks can produce similar signatures. Severe interference (profile 02) and packet loss at the SPGW (profile 03) affect throughput in a similar way while the RTT closely matches the baseline measurements. On the other hand, congestion on the SGi link (profile 08) and packet loss at the eNB (profile 03) affect both the throughput and the RTT similarly.
Further, it is clear that identical bottlenecks impact end-to-end performance in different ways depending on where they occur, as is the case with packet loss (profiles 03 and 07).  
This demonstrates how service layer measurements are not always sufficient for separating bottlenecks.

\begin{figure}[t]
	\centering
	\includegraphics[width=1\columnwidth]{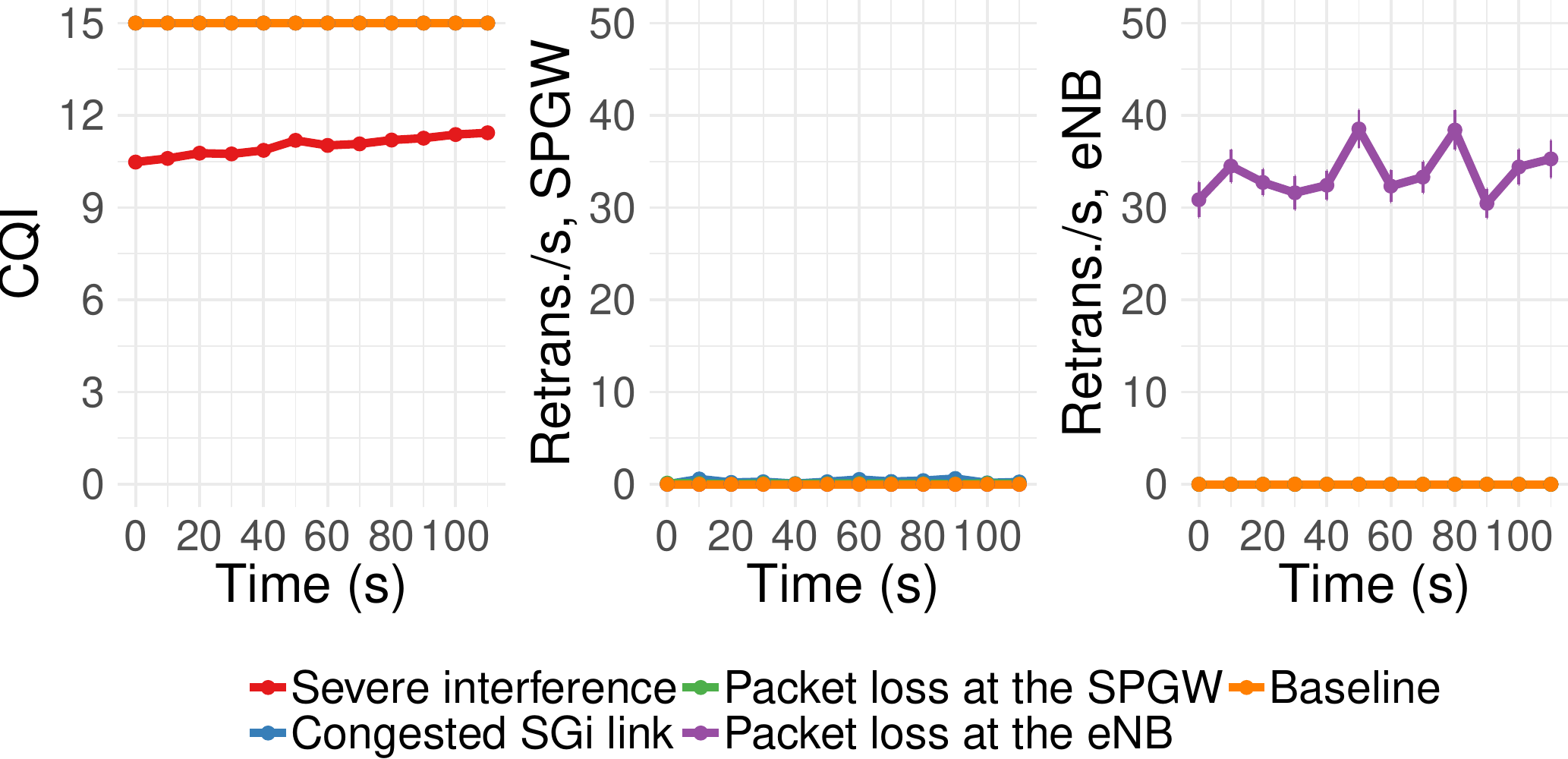}
	\vspace{-8mm}
	\caption{Network function layer measurements: CQI, TCP retransmissions on the SPGW and eNB.}
	\label{fig:third_set_nf}
	\vspace{-4mm}
\end{figure}

\noindent{\bf Network Function layer.} Factoring in NF layer measurements, Fig.~\ref{fig:third_set_nf} shows the CQI as well as TCP retransmissions at the SPGW and the eNB. 
CQI, which measures radio channel quality, helps pick out interference (profile 02), while measurements of TCP retransmissions at the eNB help isolate packet loss at the eNB (profile 03).
The remaining two profiles (profiles 07 and 08), however, are indistinguishable using NF layer measurements.
This is for two reasons: i) there are no NF layer measurements that can reliably detect link congestion and ii) the measurements for TCP packet loss can only detect retransmissions at the sender or receiver, not the intermediate hops such as the SPGW. 
The identification of TCP retransmissions in intermediate hops would require the capturing and inspection of individual users' packets which is impractical due to large associated overheads.

\noindent{\bf Infrastructure layer.}
Examination of the infrastructure layer metrics (not shown) reveals the congestion in the SGi link (profile 08) by measurements of packets transmitted and received at the SPGW, finally allowing us to separate it from packet loss (profile 07).
The remaining bottleneck of packet loss (profile 07), while evidently affects the service level performance that the user is experiencing (in this case, the throughput was severely degraded), can't be traced to a root cause through visual inspection.

\medskip

For completeness, we visually inspected all measurements over the range of bottleneck profiles to determine their utility for detecting the considered bottleneck types.
Table~\ref{table:1} distills our observations.
We grade the measurements and their relationship to specific bottlenecks as follows:
measurements that are always affected in the presence of a bottleneck (\checkmark), measurements that can be affected in the presence of a bottleneck (\checkmark ?) or measurements that are not affected in the presence of a bottleneck.
In an ideal case, each bottleneck would uniquely affect one or more measurements, making its identification simple. 
However, that is rarely the case even when disregarding the location and severity of a profile and only looking at the basic bottleneck.

As shown by the analysis above, service layer measurements that capture the end-to-end performance can serve as an indicator of a bottleneck but cannot by themselves isolate the underlying root causes.
Because all bottlenecks tend to affect them, inferring which is to blame becomes infeasible.
At the Network Function layer, the radio TX and MCS are closely linked to throughput and provide similarly ambiguous insights.
On the other hand, the Network Function layer provides a number of measurements that can be clearly linked to specific bottlenecks. Radio retransmissions and CQI are linked to interference, while TCP retransmissions are linked to packet loss.
However, as mentioned in Section~\ref{sec:methodology_bottlenecks}, packet loss may itself be caused by congestion. 
Bottlenecks of congestion, insufficient computational resources and delay cannot be clearly identified by the network function layer even though several of the measurements can provide indications of their existence.
Finally, from the infrastructure layer measurements, CPU, memory and storage utilization directly reveal computational resource bottlenecks while also providing hints on congestion due to the increased computational load that it creates. Network TX and RX reveal congestion while active measurements for each link identify delay.
Overall, although each bottleneck clearly impacts a number of specific measurements, the value of said measurements is obscured by the fact that other bottlenecks can affect them as well.

\begin{savenotes}
\begin{table}
\small
\begin{tabular}{ | m{1.5cm} | M{0.9cm} | M{0.9cm} | M{0.9cm} | M{1cm} | M{0.9cm} | }
 \cline{2-6}
    \multicolumn{1}{c|}{}
 & \textbf{\rotatebox[origin=c]{60}{Interference}} & \textbf{\rotatebox[origin=c]{60}{Packet loss}} & \textbf{\rotatebox[origin=c]{60}{Congestion}} & \textbf{\rotatebox[origin=c]{60}{Resources~\footnote{Corresponds to Insufficient Computational Resources.}}} & \textbf{\rotatebox[origin=c]{60}{Delay          }} \\
\hline
\rowcolor[HTML]{fc8d59}
Throughput & \checkmark ? & \checkmark ? & \checkmark ? & \checkmark ? & \checkmark ? \\
\hline
\rowcolor[HTML]{fc8d59}
RTT & \checkmark ? & \checkmark ? & \checkmark ? & \checkmark ? & \checkmark ? \\
\hline
\rowcolor[HTML]{ffffbf}
Radio TX & \checkmark ? & \checkmark ? & \checkmark ? & \checkmark ? & \checkmark ? \\
\hline
\rowcolor[HTML]{ffffbf}
Radio RTX & \checkmark & & & \checkmark ? & \checkmark ?  \\
\hline
\rowcolor[HTML]{ffffbf}
CQI & \checkmark & & & &  \\
\hline
\rowcolor[HTML]{ffffbf}
MCS & \checkmark ? & \checkmark ? & \checkmark ? & \checkmark ? & \checkmark ?  \\
\hline
\rowcolor[HTML]{ffffbf}
MSD & & \checkmark ? & \checkmark ? & \checkmark ? & \checkmark ?  \\
\hline
\rowcolor[HTML]{ffffbf}
TCP RTX  & & \checkmark & \checkmark ? & &  \\
\hline
\rowcolor[HTML]{91bfdb}
CPU & & &  \checkmark ? & \checkmark &  \\
\hline
\rowcolor[HTML]{91bfdb}
Memory & & &  \checkmark ? & \checkmark &  \\
\hline
\rowcolor[HTML]{91bfdb}
Storage & & &  \checkmark ? & \checkmark &  \\
\hline
\rowcolor[HTML]{91bfdb}
TX/RX  & & & \checkmark & &  \\
\hline
\rowcolor[HTML]{91bfdb}
Link Delay  & & & & & \checkmark \\
\hline
\end{tabular}
\vspace{1mm}
\caption{The measurement parameters and what they may indicate. Colors denote the system layers: service, network function and infrastructure. Check-marks indicate a direct relation while question marks indicate possible relation or relation for only some of the profiles. Empty cells indicate no discernible relation.}
\vspace{-8mm}
\label{table:1}
\end{table}
\end{savenotes}

{\em \textbf{Takeaway:} While bottlenecks may be caused by unrelated issues and originate from different points of the network, they can produce seemingly similar effects in terms of observed KPIs making their identification a challenging task.
Manual inspection of said KPIs can provide insights and narrow down the potential causes of bottlenecks but the number of sources and volume of monitoring data as well as the complexity of their mutual correlations make this approach impractical if not impossible for a production network.}

\subsection{Machine Learning based Measurement Analysis}

While visual inspection can help in attributing bottlenecks to links or VNFs, the effort required to do so quickly becomes intractable due to the sheer volume of data.
To contextualize this, the exercise in the previous subsection, examined five measurements out of the initial fifty-two.
The runs of each profile were aggregated by averaging and the temporal granularity was reduced by an order of magnitude.
Considering that our testbed consists of just seven monitored nodes including all VNFs, host machines and UEs, it is clear that visual inspection does not scale for a typical mobile network with hundreds if not thousands of RAN nodes and tens of more nodes in the CN.  
An alternative is to automate troubleshooting using machine learning (ML) based methods. 

There are two major types of ML techniques pertinent to the use-case at hand: \textbf{supervised or unsupervised learning}.
The former requires training based on known bottlenecks, which presents a number of challenges when applied to bottleneck detection in a production network.
As is the case with current threshold based alarm systems, supervised learning is based on expert analysis of data and labelling of bottlenecks for training.
The inherent variability and volatility of operational networks in terms of infrastructure, architecture and use-cases, makes supervised training complex and imposes a recurring overhead for re-training.

With little or no training data, unsupervised ML can aid our understanding of the network and development of solutions that cater to the underlying topology, user patterns and available telemetry.
It is also equally useful in the process of determining how the telemetry should be aggregated, collected and analyzed.
This is in line with our motivation to understand bottleneck identification within the new domain of cloudified mobile networks and highlight the key challenges of instrumentation.
It also allows exploring the practical considerations for implementing a complete monitoring system even if more involved techniques, such as supervised learning, may eventually be needed.

We therefore turn to unsupervised learning and more specifically to clustering since we are dealing with a partitioning problem.
Essentially, the use of clustering allows us to investigate whether the bottleneck profiles are inherently separable from the baseline performance and from each other.
It also provides a mean, that is the number of clusters, to study trade-offs between the accuracy and granularity of bottleneck identification. 

\subsubsection{Clustering}
\label{sec:clustering}

As we are looking to partition our profile runs to a number of clusters, the minimum intervention needed, is to determine the number of clusters that we are searching for.
For the evaluations presented in this section, the number of clusters is set to \textit{k~=~number~of~profiles}, which is 17 in our case as outlined in Table~\ref{table:2}.
Recall that by a profile run we are referring to a bottleneck type, intensity and location. 
In this way, we are looking at a best case scenario, where the distinct bottleneck profiles match the number of clusters and we attempt to group together the runs of each profile.

\begin{table}
\small
\begin{tabular}{ | m{1.8cm} | m{0.3cm} | m{5.5cm} | } 
\hline
\textbf{Bottleneck} & \textbf{ID} & \textbf{Profile} \\
\hline
\multirow{2}{1.8cm}{Interference} 
& 01 & Moderate \\ \cline{2-3}
& 02 & High \\ \hline
\multirow{5}{1.8cm}{Packet Loss} 
& 03 & low at the SPGW, 1\%\\ \cline{2-3}
& 04 & moderate at the SPGW, 4\%\\ \cline{2-3}
& 05 & high at the SPGW, 6\%\\ \cline{2-3}
& 06 & moderate at the controller, 4\%\\ \cline{2-3}
& 07 & moderate at the eNB, 4\%\\ \hline
\multirow{3}{1.8cm}{Congestion \newline (link of)} 
& 08 & SPGW - external network \\ \cline{2-3}
& 09 & SPGW's host machine - external network \\ \cline{2-3}
& 10 & controller - eNB \\ \hline
\multirow{3}{1.8cm}{Insufficient \newline Computational \newline Resources} 
& 11 & at the SPGW (CPU/memory/storage) \\ \cline{2-3}
& 12 & at the SPGW host (CPU/memory/storage) \\ \cline{2-3}
& 13 & at the controller (CPU/memory/storage) \\ \hline
\multirow{4}{1.8cm}{Delay \newline (added at)} 
& 14 & moderate at the SPGW, 30ms \\ \cline{2-3}
& 15 & moderate at the controller, 0.9ms \\ \cline{2-3}
& 16 & moderate at the eNB, 0.9ms \\ \cline{2-3}
& 17 & high at the eNB, 1.5ms \\ \hline
\end{tabular}
\vspace{1mm}
\caption{Summary of bottleneck profiles.}
\vspace{-8mm}
\label{table:2}
\end{table}

Given that we have perfect knowledge of how our profiles should be clustered, we evaluate the performance of clustering using the {\em purity metric}~\cite{schutze2008introduction}, a measure of the extent to which clusters contain only a single class.
Accordingly, a maximum purity of 1 denotes that the profiles have been grouped together perfectly, meaning that each cluster only contains samples produced from a single bottleneck profile.
This ideal outcome would mean that given the measurements at our disposal and our clustering method, it is possible to perfectly distinguish between the profiles in an automated manner.

In the process of analyzing our data, we were faced with a multitude of choices for specific algorithms, distance metrics and feature-sets that we evaluate below.
Before delving into this vast search space, we need to appropriately describe our measurements by defining features that can readily be used by ML techniques.
Each measurement consists of a 120-second long timeseries.
For each such timeseries, the extracted features consist of the first four moments of its distribution, namely its mean, variance, skewness and kurtosis, along with the minimum and maximum values.
These features are chosen to capture the properties of underlying probability distributions.
This produces a total of 312 features extracted from 52 monitored parameters.

\subsubsection{Algorithms and Distance Metrics}
\label{sec:analysis_algorithms}

A number of clustering algorithms can partition a data-set into a known number of clusters. We evaluated the performance of three commonly used algorithms~\cite{schutze2008introduction} and their variations, namely \textit{K-means}, \textit{Agglomerative (hierarchical)} clustering and \textit{DBSCAN}.

Hierarchical clustering proved to be the most effective due to its simplicity and flexibility in terms of distance metrics and linkage criteria.
Hierarchical clustering is compatible with a score of distance metrics and a number of well known linkage criteria (i.e. the methods for grouping observations into clusters).
Distances in the Intersection and the L-1 family namely the Soergel distance (L-1) and its equivalent, the Tanimoto distance (intersection), proved most suitable in our analysis~\cite{cha2007comprehensive}.

The results presented in this study henceforth are thus based on hierarchical clustering using the Soergel/Tanimoto distance.

\subsubsection{Feature-Sets and Selection}
\label{sec:analysis_features}

\begin{figure}[t] 
	\centering
	\includegraphics[width=1\columnwidth]{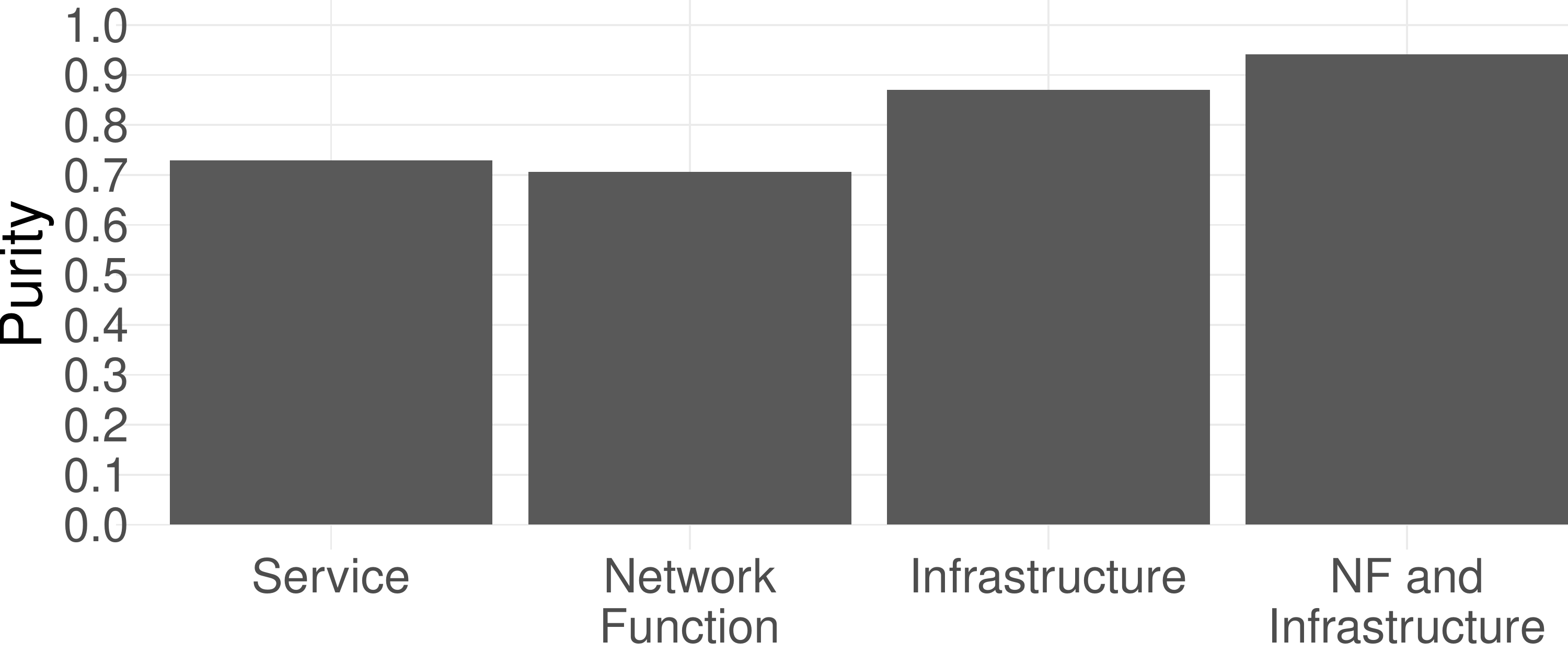}
	\vspace{-8mm}
	\caption{Comparison of purity achieved by the different feature-sets.}
	\vspace{-4mm}
	\label{fig:features}
\end{figure}

Having identified a suitable clustering algorithm, we now turn to evaluate the utility that each layer of measurements brings and whether needed measurements can be reduced while also ensuring high level discrimination among bottlenecks (as reflected by purity).
During visual inspection (Sec.~\ref{sec:analysis_initial}), we find that a number of the collected measurements do not at all deviate from the baseline.
We therefore identify and remove all such features.
These are largely made up of measurements from the control-plane VNFs of the network which remain unaffected by data-plane bottlenecks (i.e. the HSS and MME) such as high packet-loss, congestion, insufficient computational resources.
Out of an initial total of 312 features, 186 remain, representing 31 out of the 52 measurements (a 60\% reduction). 
This boosts the accuracy of clustering by reducing the noise of the data-set and focusing on the important features. 
Following feature reduction, as shown in Fig.~\ref{fig:features}, we cluster based on different subsets of features that match the layers described in Sec.~\ref{sec:analysis_initial}.
Next, we discuss the results with respect to each feature-set.

\begin{figure*}[t]
\begin{center}
    \subfigure[Service layer]
{\includegraphics[width=0.48\linewidth]{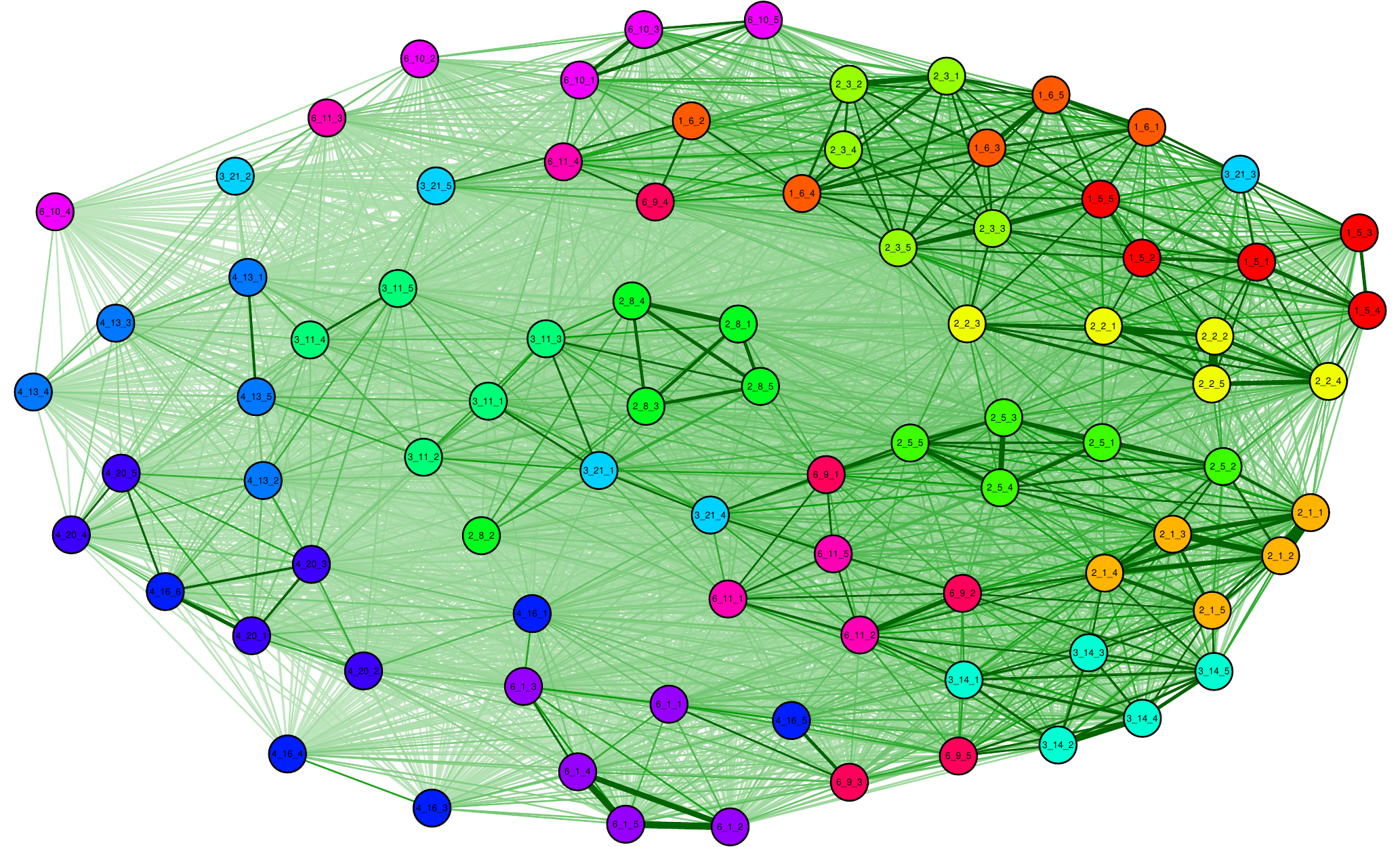}\label{fig:distance_plot_active}}
\hfill
    \subfigure[NF and Infrastructure layers]
{\includegraphics[width=0.48\linewidth]{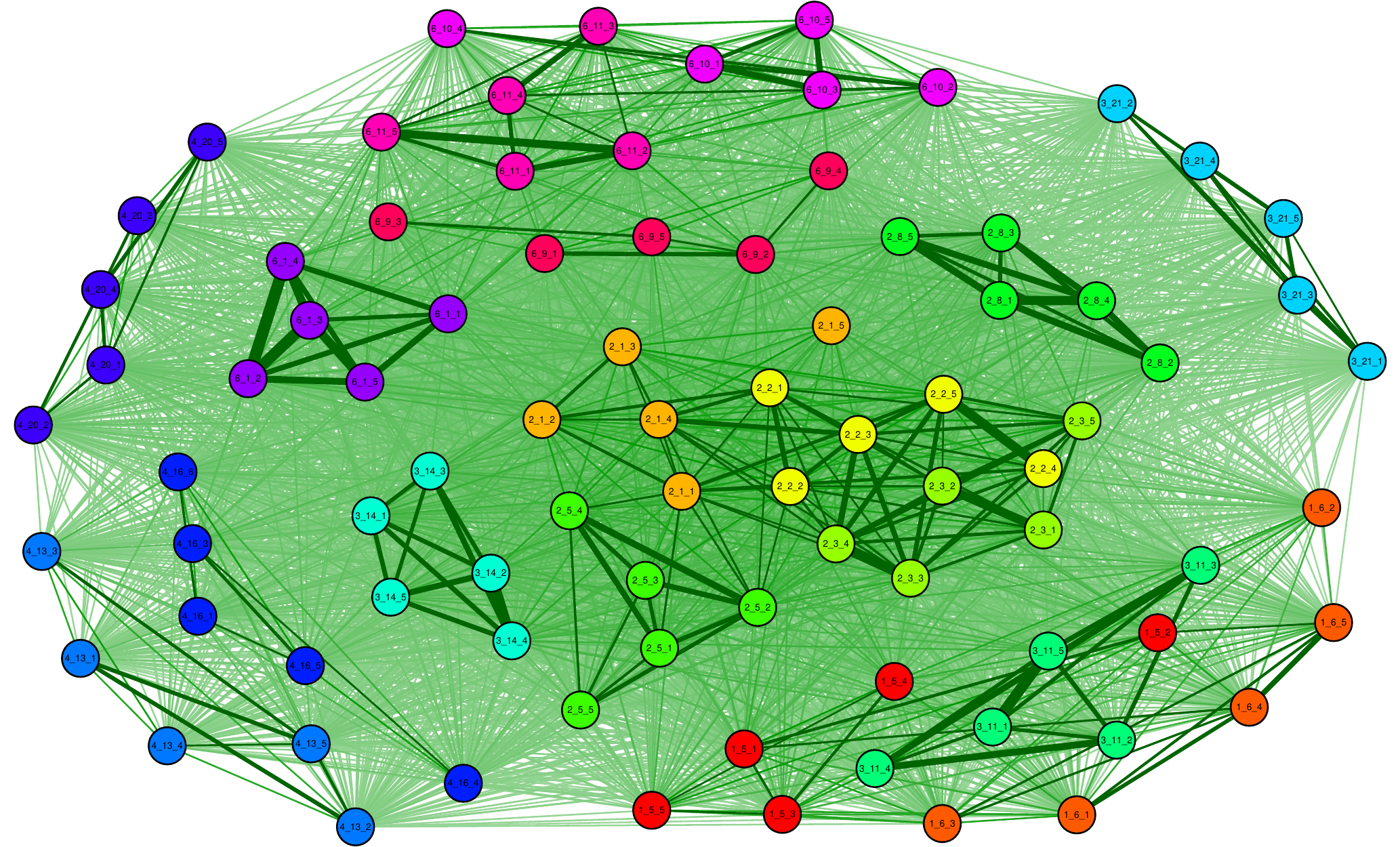}\label{fig:distance_plot_all}}
\vspace{-4mm}
\caption{The figures compare the distances produced by the service layer measurements (shown in (a)) with those produced by the NF and infrastructure layer measurements (b). While in (a) confusion of the profiles (colors) is visible, in (b) intra-profile distances are visibly shorter while inter-profile distances are larger.}
\vspace{-4mm}
\label{fig:distance_plot}
\end{center}
\end{figure*}

\textbf{Service layer} features provide a relatively low purity of $\approx 0.7$. 
This is because all bottlenecks affect throughput and RTT and in many cases different profiles have a very similar impact, thus rendering these metrics suitable only as bottleneck indicators and not as discriminators.
To better visualize the confusion that comes from using the service layer features alone, Fig.~\ref{fig:distance_plot_active} places our data-set in two-dimensional space with each profile represented by a color and its profile's runs represented by a point.

\textbf{Network Function layer} features do not yield better results.
Thanks to the availability of information about the channel quality, interference is now identifiable as are some of the packet loss profile runs, thanks to the measurements of TCP retransmissions.
However, the profiles introducing congestion, overload and delay as well as the profiles of packet loss in the CN are not easily distinguishable due to the lack of NF layer measurements that directly identify them.

The \textbf{Infrastructure layer} provides the largest feature-set, including measurements of the network and computational resources that can uncover traffic patterns, delay and computational hot-spots throughout the network.
As a result, all of the profiles introducing congestion, insufficient computational resources and delay are successfully identified.
Due to the lack of information about the channel quality, interference cannot be reliably picked out.
Finally, due to lack of measurements to detect packet loss in intermediate hops, confusion remains within the various packet-loss profiles.

\textbf{Network Function and Infrastructure.}
After separately examining each layer, we combine measurements from the NF and infrastructure layers.
As, we have seen above, as opposed to the service layer measurements, these two layers bring a number of measurements that could help separate various bottlenecks.   
The combined set can identify all of the profiles except for some packet loss related ones due to the absence of packet loss measurements at intermediate hops.  
To better visualize the improvement in intra-profile and inter-profile distances, Fig.~\ref{fig:distance_plot_all} places our data-set in two-dimensional space, with the distances calculated by the combined feature-set across NF and infrastructure layers.

\begin{figure}[t]
	\centering
	\includegraphics[width=1\columnwidth]{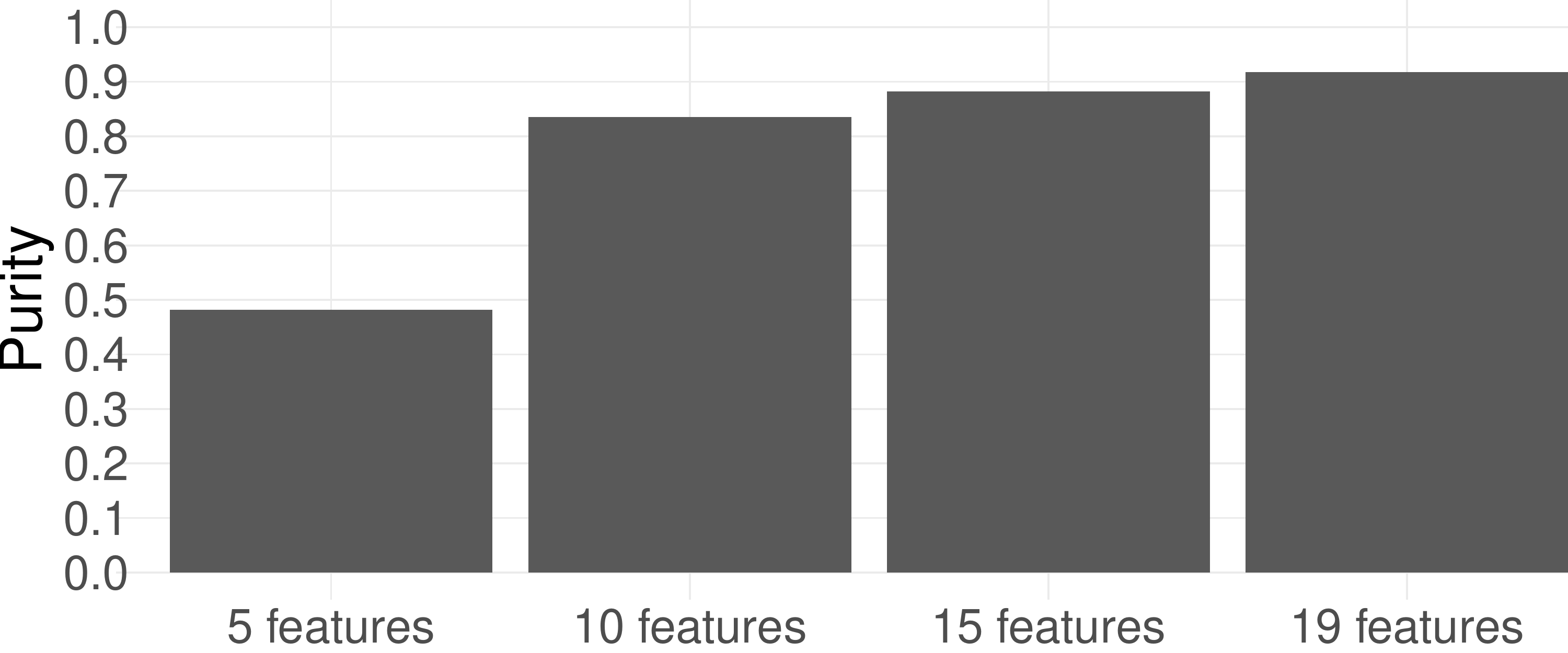}
	\vspace{-8mm}
	\caption{Comparison of purity achieved by the number of selected features.}
	\label{fig:selection}
	\vspace{-4mm}
\end{figure}

\textbf{Feature selection.}
In addition to the layer based feature-sets we examine whether a smaller subset of features can be constructed that can provide accuracy comparable to our best results.
We use non-parametric regression and specifically the Multivariate Adaptive Regression Splines technique~\cite{friedman1991multivariate} to build a new feature-set based on the most important features.
Fig.~\ref{fig:selection} shows that just 10 features result in a purity exceeding that of the service layer features and a mere 19 features can achieve purity comparable to that of the combined feature-set across NF and infrastructure layers.
These 19 features represent 13 out of the 31 measurements, a further 59\% reduction following feature reduction. 
This demonstrates that there is potential to significantly reduce the overhead of measurement collection and analysis while maintaining accuracy.
The 19 selected features are based on measurements of:
MCS,
delay of the controller-eNB and S1-U links,
TCP errors and memory utilization of the controller,
bytes transmitted and memory utilization of the SPGW,
bytes transmitted of the SPGW's host,
TCP errors of the UE,
bytes received and transmitted, TCP errors and memory utilization of the eNB.
The features intuitively cover link delay, and utilization metrics that can directly indicate three types of bottlenecks: link congestion, insufficient computational resources and delay.
The service layer is represented by the MCS and TCP retransmissions. Directly indicating packet loss and indirectly indicating interference.

{\em \textbf{Takeaway:} Commonly used ML algorithms when properly tuned can greatly ease the process of analyzing measurements, aiding in discriminating between bottlenecks. Combining measurements from different layers greatly improves the potential for separating bottlenecks.  
Furthermore, feature selection significantly reduces the feature-set while trading off little accuracy.
The selected features come from different layers and locations of the network, allowing detection of all the bottlenecks tested and emphasizing the importance of a diverse set of measurements.}

\section{Automating bottleneck identification and attribution}
\label{sec:bottlenecks}

We have shown in the previous section that precisely knowing the exact number of profiles, we are able to separate them. Next, we explore the practical trade-offs pertinent to bottleneck identification from a network monitoring perspective.
Recall that in our setup we have 17 bottleneck profiles, where each profile is defined by bottleneck type, location (i.e. which VNF or link) and intensity (i.e. how severe).
Predetermining the number of bottleneck profiles that may occur in a commercial network with thousands of nodes, users, traffic types and patterns is a daunting if not an impossible task.
With that in mind, a system designed to automatically identify performance bottlenecks will need to be based on a coarse-grained categorization.
Hence, we need to lay out a logical categorization that covers a majority of potential bottlenecks and do so in a way that will provide information to sufficiently identify and localize them.

Here, we present an evaluation of bottleneck identification based on two such categorizations: per bottleneck type; and per bottleneck type and location.
In other words, starting from the complete set of profiles (Table~\ref{table:2}), the first categorization abstracts away bottleneck location and severity, while the second abstracts away bottleneck severity.
To achieve this, we explore two approaches.
The first approach is based on a centralized methodology where a global view is available for all measurements in the network. The second approach is based on a distributed methodology where analysis is undertaken at several places requiring only a local (i.e. partial) view of the network's measurements.
We evaluate them using the four layer based feature-sets plus one based on selection of the most important parameters as in the previous section.

\subsection{Centralized Approach}
\label{sec:bottlenecks_1step}

Going by the centralized methodology, a monitoring system will retrieve measurements from several elements across the network and collect them to a central point of analysis.
Without considering the disadvantages of this approach regarding the delay and overhead of shipping the measurements across the network, it undoubtedly provides the advantage of a global and complete awareness of the network state.

\subsubsection{Categorization by bottleneck type}
\label{sec:bottlenecks_1step5}

\begin{figure}[t]
	\centering
	\includegraphics[width=1\columnwidth]{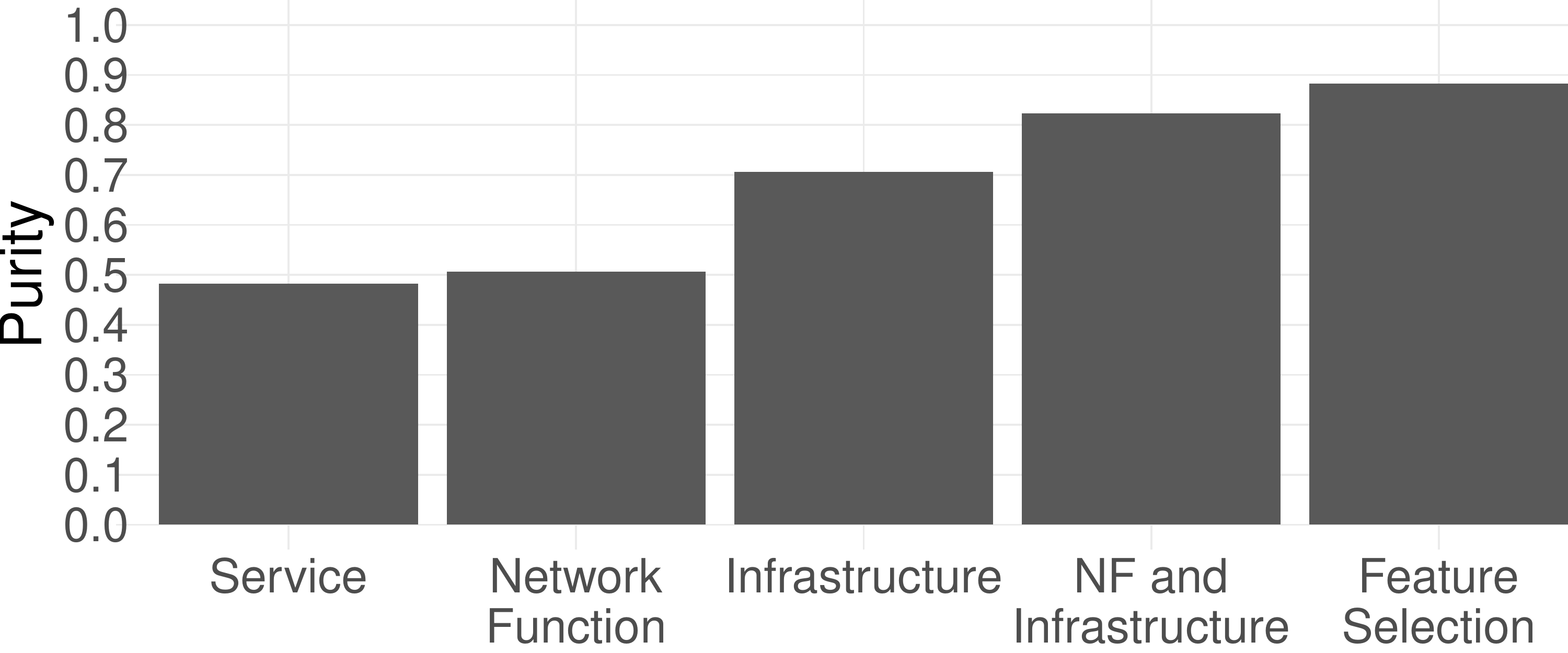}
	\vspace{-6mm}
	\caption{Bottleneck type identification.}
	\vspace{-4mm}
	\label{fig:1step_5clusters}
\end{figure}

We cluster the profiles according to the basic five types of bottlenecks that may impact a mobile network as described in Sec.~\ref{sec:methodology_bottlenecks}.
Fig.~\ref{fig:1step_5clusters} shows the purity obtained when clustering into these five types using different sets of features.

In line with our earlier observations in Sec.~\ref{sec:analysis_features}, both service layer and network function layer features perform poorly because they lacking features necessary for identifying infrastructure-related bottlenecks.
The purity hikes to 0.71 when using the infrastructure features.
However, the lack of features regarding radio link performance and packet loss in intermediate hops, leads to confusion of interference and packet loss as well as confusion of packet loss with delay.

Combining the network function and infrastructure features increases purity further to 0.82 as this set allows for separating RAN as well as CN problems.
However, confusion remains amongst profiles of packet loss and others like congestion because we lack features that capture loss on the path.

The feature-set based on feature selection provides the highest purity of 0.88, by eliminating measurements that were contributing to confusion of edge cases, while using 65\% less measurements, from 31 down to 11.

In total, 13 features are selected based on the measurements of CQI, MCS, delay of the controller-eNB and S1-U links, bytes transmitted and memory utilization of the controller and SPGW, bytes received and memory utilization of the SPGW's host, bytes received and memory utilization of the eNB.
Nevertheless, once again both the NF and infrastructure layer are represented with measurements that can conceptually identify all of the bottlenecks.

\subsubsection{Categorization by bottleneck type and location}
\label{sec:bottlenecks_1step9}

\begin{figure}[t]
	\centering
	\includegraphics[width=1\columnwidth]{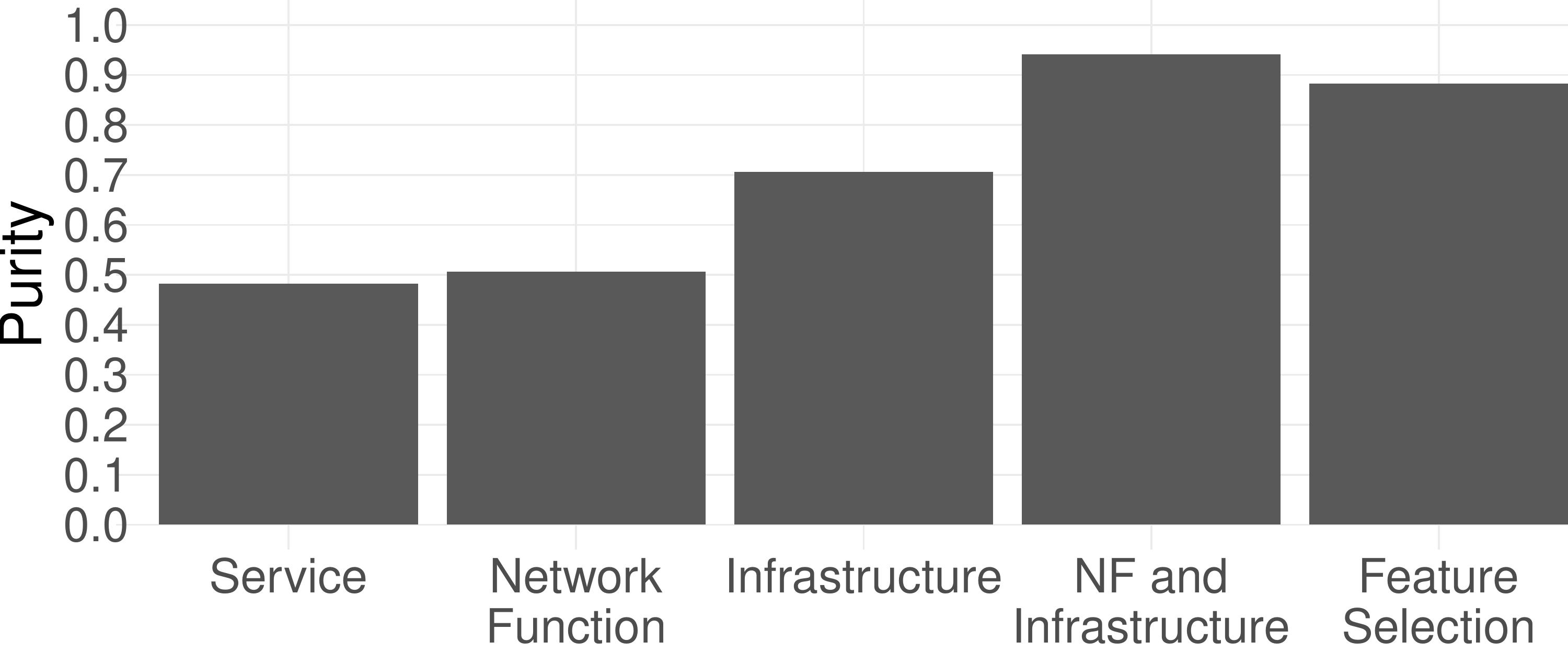}
	\vspace{-8mm}
	\caption{Bottleneck type and location identification.}
	\vspace{-4mm}
	\label{fig:1step_9clusters}
\end{figure}

While identifying the type of bottleneck is certainly a good start, knowing where the problem lies is equally valuable.
Here we move to a more fine-grained categorization where we match bottlenecks to nine clusters, according to both their type and location.
When considering location, we aggregate the components of the network into two major groups, as dictated by the network architecture. The eNB and controller are part of the RAN, while the HSS, MME, and SPGW form the CN. Aggregating the components based on their respective domain allows us to determine whether the information captured by the measurements can provide the dimension of location through automated analysis. This approach should be seen as a first step towards advanced analytics that would seek to pinpoint a bottleneck's exact location i.e. at the level of a function or a link.
In total, we have five possible bottlenecks in the RAN (i.e. those described in Sec.~\ref{sec:methodology_bottlenecks}) and four in the CN (same as the RAN except interference), which add up to nine clusters.
Fig.~\ref{fig:1step_9clusters} illustrates the purity for the different feature-sets when clustering our data into nine clusters.

Using the service layer, network function layer or infrastructure layer features alone results in purity similar to what we previously measured (see Fig.~\ref{fig:1step_5clusters}).
Increasing the number of clusters does not result in better purity due to the availability of only a single layer of measurements that cannot identify all 5 bottlenecks in each location.

\begin{figure}[t]
	\centering
	\includegraphics[width=1\columnwidth]{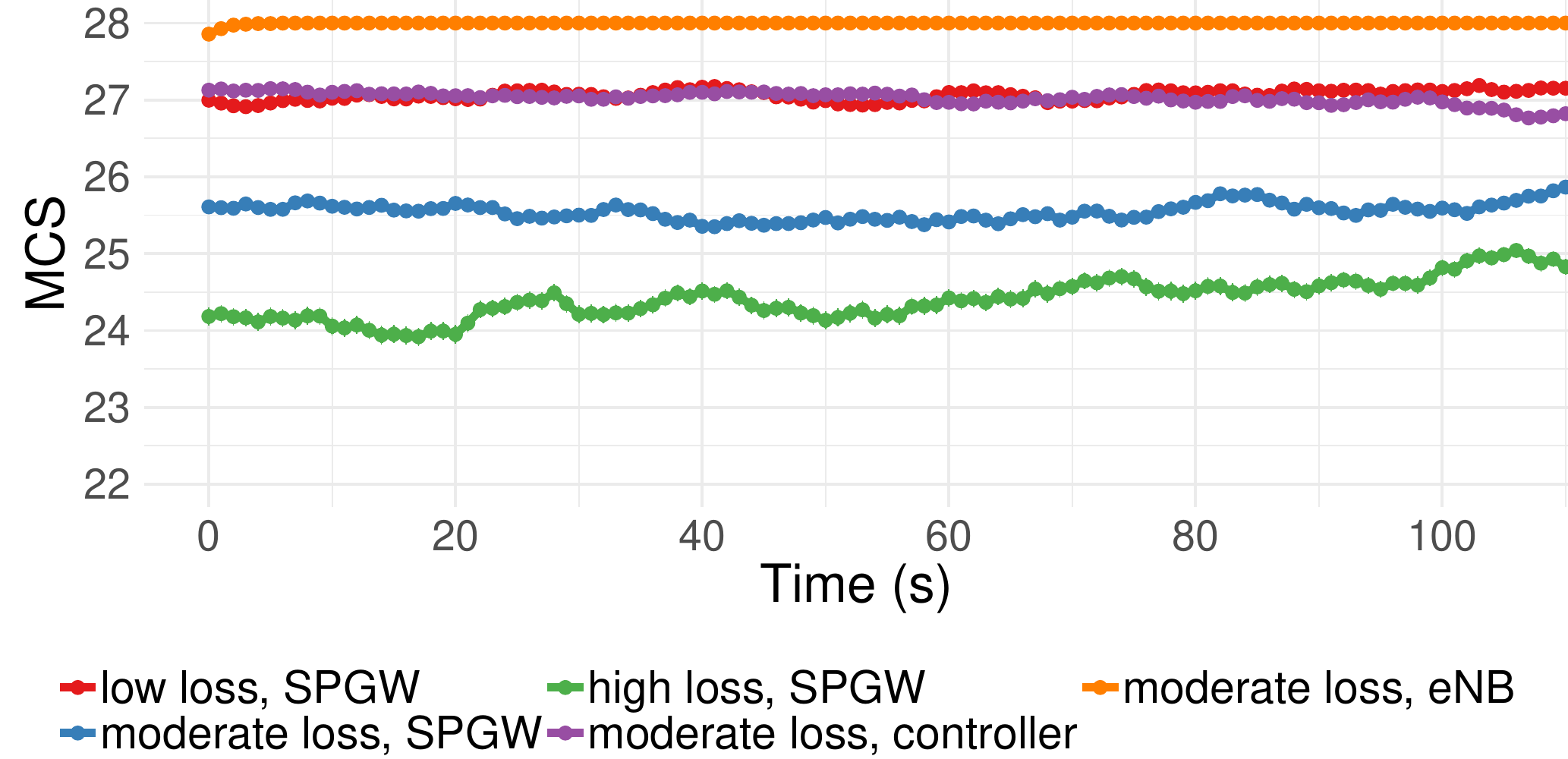}
	\vspace{-8mm}
	\caption{Packet loss in the controller or the CN affects the MCS similarly. On the other hand, packet loss in the eNB provides very stable measurements.}
	\vspace{-4mm}
	\label{fig:fourth_set_mcs}
\end{figure}

The combined feature-set provides excellent results with a purity of 0.94.
The only remaining source of confusion is packet loss.
As an example, packet loss in the RAN controller is misclassified as packet loss in the CN instead of packet loss in the RAN.
This is because we lack per-hop measurements of packet loss.
Reverting to metrics that capture the end-to-end performance like throughput or looking at MCS and MSD, which can potentially indicate packet loss (see Table~\ref{table:1}), does not help either, as illustrated by the example of visual inspection of the MCS in Fig.~\ref{fig:fourth_set_mcs}.
Once again, the problem with metrics that can seemingly capture a wide range of bottlenecks is exemplified.
While the MCS provides easy differentiation among levels of severity in the SPGW, once we introduce bottlenecks at multiple locations the results are obscured and no longer dependent on the severity of packet loss (or the type of bottleneck altogether).
Furthermore, aggregate network counters like the number of sent and received packets will only help if packet loss is severe.

Feature selection provides slightly lower purity at 0.88  while using 65\% less measurements. It introduces confusion between delay on the S1-U link and insufficient computational resources on the controller. 
In addition to the MCS, this is due to the missing measurements on the state of CPU and storage load on the controller. 
Also, packet loss on the controller and packet loss on the eNB mistakenly created two separate clusters.
The features selected here are based on measurements of CQI, MCS, missed scheduling deadlines, delay of the controller-eNB and S1-U links, TCP retransmissions and memory utilization of the controller, bytes received and memory utilization the SPGW's host, TCP retransmissions and memory utilization of the eNB. 
Once again, both the network function and infrastructure layer are represented with measurements that can conceptually identify all bottlenecks.

{\em \textbf{Takeaway:} Interestingly, a coarse grained segregation based on bottleneck type can be reasonably accurate provided we choose the right set of features.
Introducing the dimension of location allows for a fine grained categorization of bottleneck profiles. 
This gives deeper insights that can be exploited for successfully troubleshooting the bottleneck. Furthermore, combining the right measurements from various layers is crucial for boosting the accuracy of bottleneck identification.}

\subsection{Distributed Approach}
\label{sec:bottlenecks_2step}

With a distributed methodology, a monitoring system provides multiple points of analysis where measurements retrieved from the local network elements can be collected.

Here we consider an implementation with two analysis points: one to handle measurements local to the RAN and another one to handle those local to the CN.
This approach essentially separates RAN and CN bottlenecks in one step and then identifies the respective type of bottleneck within each domain in a second step.
Once again, this should be seen as an exploratory step towards a full featured monitoring system that could conceivably be distributed further, even to the point of individual VNFs.
Such distribution has the potential to greatly reduce the delay and overhead of shipping the measurements across the network but it faces the challenge of incomplete awareness of the network state.

Applying this to our measurements we achieve perfect purity during the first step, when using the combined feature-set or feature-selection.
This means that both the RAN and the CN analysis points successfully identify profiles that affect their domain, while ignoring the specific type of bottleneck.
For the second step, clustering is performed anew at the respective domain with a new objective, to obtain the type of bottleneck.
This procedure finally provides the five types of bottlenecks present in the RAN and four bottlenecks in the CN (interference is not applicable) for a total of nine clusters as with the centralized approach of Sec.~\ref{sec:bottlenecks_1step9}.

\begin{figure}[t]
	\centering
	\includegraphics[width=1\columnwidth]{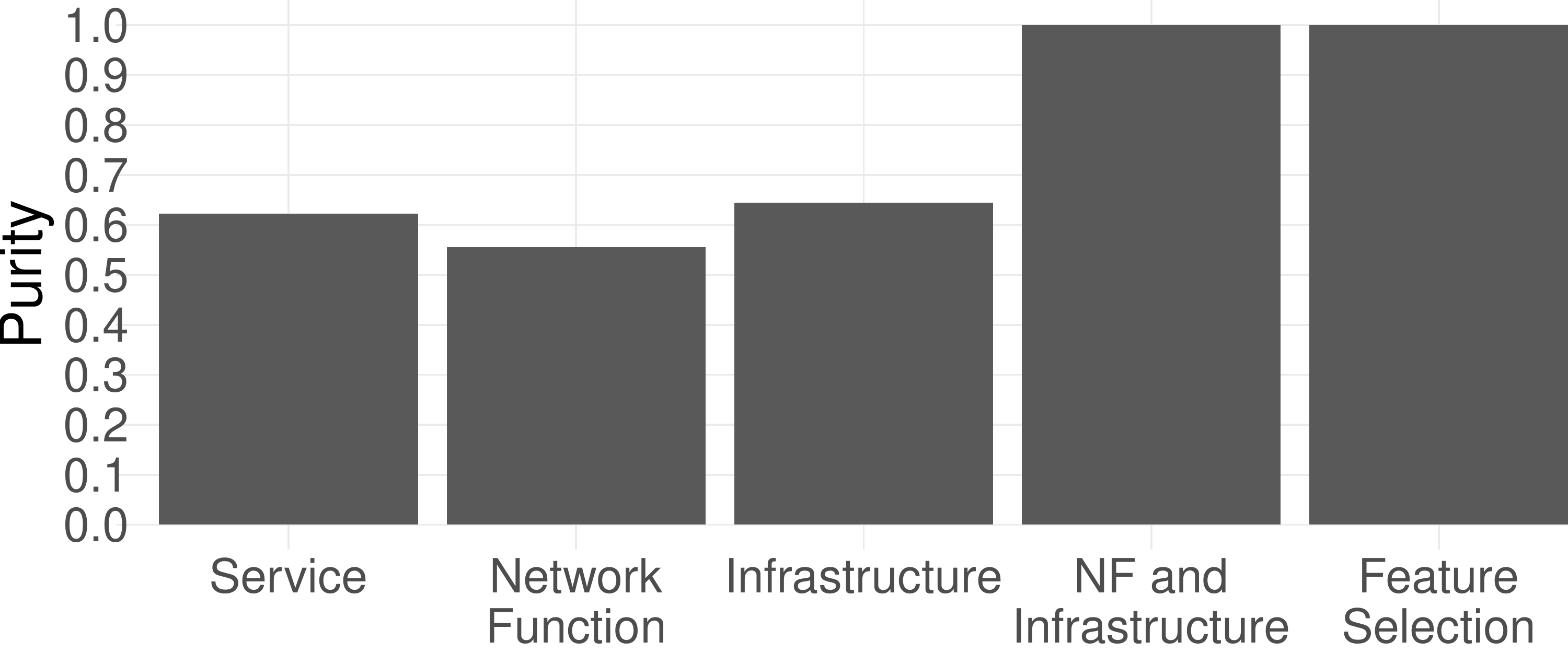}
	\vspace{-8mm}
	\caption{Bottleneck type identification in the RAN}
	\vspace{-4mm}
	\label{fig:2step_5clusters}
\end{figure}

Fig.~\ref{fig:2step_5clusters} shows the purity we obtain in the RAN, with the various feature-sets.
Once again, the service layer as well as the network function layer measurements are insufficient to provide a meaningful categorization of the profiles. However, the combined feature-set achieves perfect purity as does the selected subset of features using 53\% fewer measurements, from 19 measurements that the RAN provides to 9.

\begin{figure}[t]
	\centering
	\includegraphics[width=1\columnwidth]{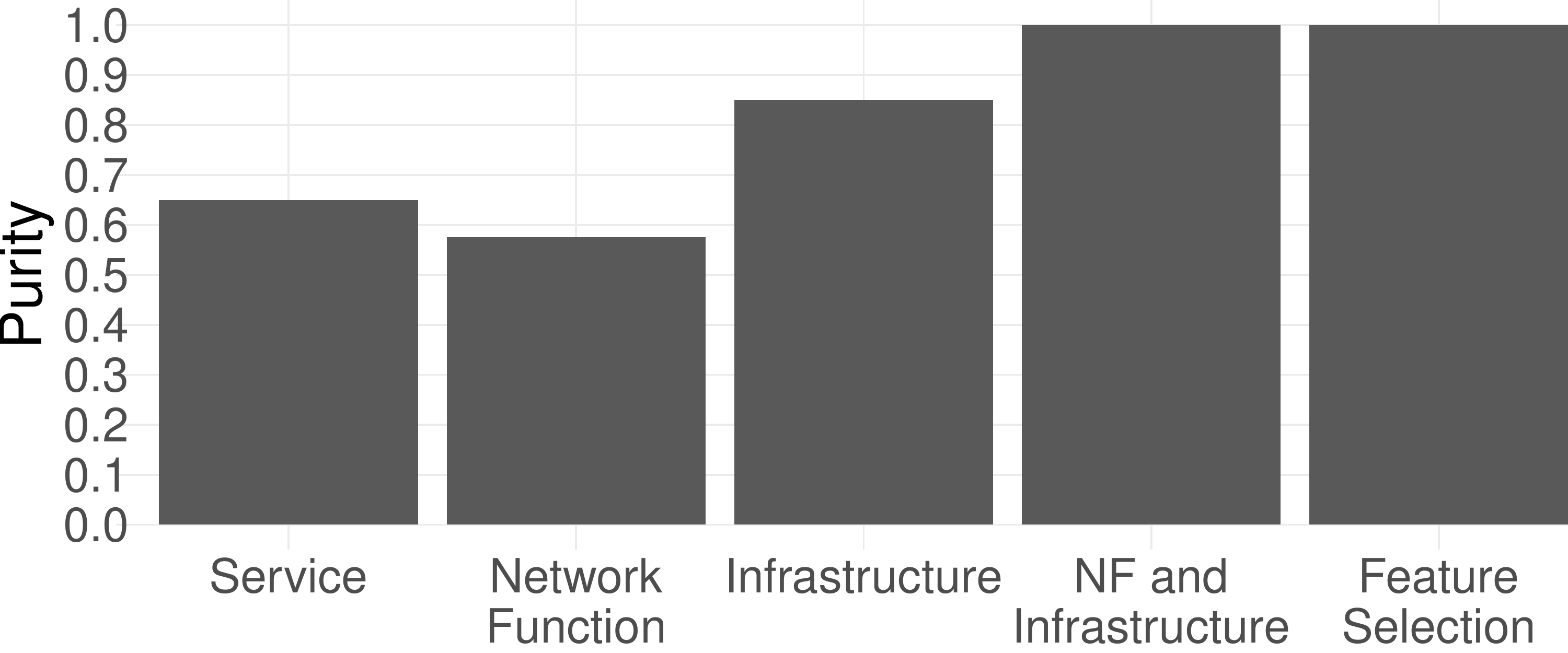}
	\vspace{-8mm}
	\caption{Bottleneck type identification in the CN}
	\vspace{-4mm}
	\label{fig:2step_4clusters}
\end{figure}

In the CN, as shown in Fig.~\ref{fig:2step_4clusters}, the outcome is similar. This time, as no interference is present, the infrastructure layer achieves higher purity of 0.85, only confusing a congestion profile for insufficient computational resources.
Finally, the combined feature-set achieves perfect purity as does the selected subset of features using 75\% fewer measurements, from 12 measurements that the CN provides to 3.

{\em \textbf{Takeaway:} Distributing the task of bottleneck identification yields higher accuracy when compared to the centralized approach. This is a consequence of the distributed analytics using only those measurements that are relevant to bottlenecks in its domain, reducing unnecessary noise.
In addition, the process can be performed concurrently with the identification of RAN problems performed at the RAN side, while the identification of CN problems is performed at the CN side.
This has the potential to greatly cut down on data-transfer overhead and identification latency, provided that analytics capabilities can be provisioned at both locations, something we explore further in Sec.~\ref{sec:discussion_overhead}.}
\section{Composite bottlenecks}
\label{sec:composites}

Bottlenecks within a network will not always manifest in isolation. 
Each user is connected to the external network through a chain of VNFs, either directly in the data path (e.g. the eNB and SPGW) or as part of the control plane of the network (e.g. the MME and HSS). 
Any part of the VNF chain, the infrastructure that it runs on and the links that connect it may experience a bottleneck. 
In addition, bottlenecks will often propagate to several locations e.g. temporal user-traffic patterns that create congestion throughout the network, or trigger cascading problems e.g. interference causing retransmissions and by extension, delay and packet loss.
In fact, these kinds of complex problems are expected to make up the majority of bottlenecks in an operational network.
In this section, we evaluate how our methodologies cope with such composite bottlenecks.

We introduce new profiles that cover three distinct composite bottleneck categories: a) same type of bottleneck at different locations; b) different bottlenecks at the same location; c) different bottlenecks at different locations. These are created from entirely new runs, where two individual bottlenecks, previously examined in sections \ref{sec:analysis} and \ref{sec:bottlenecks}, manifest at the same time as outlined in Table \ref{table:3}.

\begin{table}
\small
\begin{tabular}{ | m{2.7cm} | m{0.4cm} | m{4.4cm} | } 
\hline
\textbf{Bottleneck} & \textbf{ID} & \textbf{Profile} \\
\hline
\multirow{2}{2.7cm}{Same bottleneck \newline Different locations} & \multirow{2}{0.4cm}{18}
& Congestion (ID 8) \newline SPGW - external network \\ \cline{3-3}
& & Congestion (ID 10) \newline Controller - eNB \\ \hline
\multirow{2}{2.7cm}{Different bottlenecks \newline Same location} & \multirow{2}{0.4cm}{19}
& Resources (ID 13) \newline Controller (CPU/memory/storage) \\ \cline{3-3}
& & Delay (ID 15) \newline Moderate at the controller, 0.9ms \\ \hline
\multirow{2}{2.7cm}{Different bottlenecks \newline Different locations} & \multirow{2}{0.4cm}{20}
& Loss (ID 03) \newline Low at the SPGW, 1\% \\ \cline{3-3}
& & Delay (ID 15) \newline Moderate at the controller, 0.9ms \\ \hline
\end{tabular}
\vspace{1mm}
\caption{Summary of Composite bottleneck profiles.}
\vspace{-8mm}
\label{table:3}
\end{table}

\begin{figure}[t] 
	\centering
	\includegraphics[width=1\columnwidth]{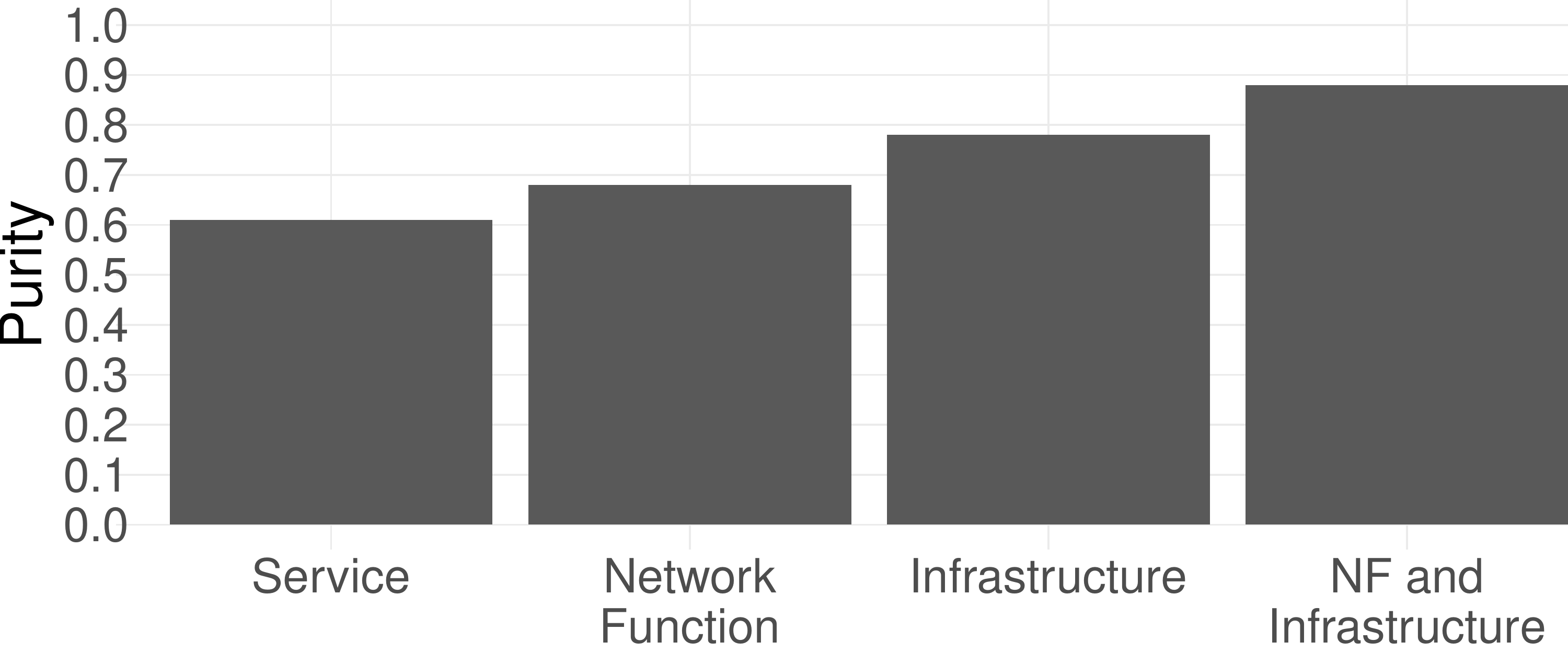}
	\vspace{-8mm}
	\caption{Comparison of purity achieved by the different feature-sets.}
	\vspace{-4mm}
	\label{fig:features_composite}
\end{figure}

At this point, we need to re-consider our options for assigning profiles to a cluster. One option would be to set the number of clusters to the number of bottleneck profiles, assuming a priori knowledge. 
Recall from Sec.~\ref{sec:analysis} that we initially have 17 profiles, adding the three new composite profiles increases this number to 20. 
Setting the number of clusters to 20, the purity that we can achieve is degraded for single layer feature-sets as seen in Fig. \ref{fig:features_composite}.
This is expected as the measurements provided by the individual layers can't sufficiently distinguish bottlenecks that span multiple layers (e.g. profile 19).
However, when using the feature-set combining measurements from the network function and infrastructure layers, we achieve purity similar to the original 17 profiles.
The composite profiles are successfully identified showing that the information captured by our complete suite of measurements can sufficiently describe the bottlenecks.

While interesting, this naive approach is clearly impractical as the number of clusters would need to scale with the number of possible bottleneck combinations which quickly becomes intractable, increasing exponentially with the number of VNFs, computing infrastructure, network and radio links etc..

Moving on to the options explored in Section~\ref{sec:bottlenecks} we use the \textbf{centralized approach} to fit the composite profiles within the 9 previously defined categories (Sec.~\ref{sec:bottlenecks_1step9}).
While not logically valid as the composite bottlenecks would belong to two clusters at the same time, this provides us with some interesting  observations.
Profile 18 is identified as one of its composites, congestion in the RAN, while congestion in the CN goes undetected.
Profile 19 is again identified as one of its composites, stress on the controller, while the delay present on the controller goes undetected. In addition, introduction of this profile causes additional confusion among previously identified experiments, lowering overall purity.
Finally, profile 20 is once again identified as one of its composites, loss at the SPGW, while the delay present on the controller goes undetected.
In every case there is no apparent ambiguity, with each run of the composite profiles placed in the same cluster. 
These bottlenecks appear as if they are simple in nature, with one of their components completely ignored.

The failure of the centralized approach in attributing composite bottlenecks to their constituents could be related to the rigid and compact nature of the used clustering approaches.    
We therefore try \textbf{fuzzy clustering} as an alternative to the strict clustering approach that we have explored so far.
Unlike hierarchical clustering, with fuzzy techniques data-points, or in this case a bottleneck, can effectively belong to multiple clusters at the same time.
To explore this we used c-means\cite{bezdek1984fcm}, a widely used fuzzy classification algorithm that allows us to predefine the number of clusters and is capable of accepting the distance metrics established in \ref{sec:analysis_algorithms}.
In much the same way as K-means, the algorithm assigns data-points to each cluster based on their distances to the cluster centroid.
However, in the case of c-means, each point is assigned a grade of membership to each cluster, making it possible for a profile to be a member of multiple clusters in varying degrees.
Profile 18 is identified as one of its composites, seen as congestion in the RAN, while congestion in the CN goes undetected.
Profile 19 is partially identified as stress on the controller with 60\% certainty, however the delay component is not identified and confused with congestion.
Profile 20 is partially identified as both of its components.
However, it is linked to the CN packet loss with a higher likelihood, that is 70\%.
Overall, the results were mixed with no composite profile being completely identified.
Further, with the fuzzy approach to clustering the accuracy of individual bottleneck identification degraded. 
However, on a positive note, except for profile 18, fuzzy clustering acknowledges that composite bottlenecks have more than a single constituent.

The lackluster performance of the centralized and fuzzy clustering approaches above suggests that an architecture-aware approach could help delineating causes of bottlenecks that span several network components.      
Next, we attempt to identify the bottlenecks using a \textbf{distributed approach}.
Here, the separate points of analysis in the RAN and CN can independently detect bottlenecks affecting their domain. 
In practice, this means that each point of analysis has a higher probability of detecting a relevant part of a composite bottleneck even though it may not reveal its full extent.
In the case of profile 18 where a single bottleneck manifests in two locations, one in each part of the network, both locations are correctly identified by the corresponding point of analysis.
By combining this information we can accurately detect the complete composite bottleneck.
In the case of profile 19, both bottlenecks are located on the RAN side. This leads analytics on the CN side to incorrectly identify the bottleneck as loss at the SPGW. On the other hand, analytics on the RAN side, behave similarly to the centralized approach by correctly detecting one of the components of the bottleneck, stress on the controller.
Finally, in the case of profile 20, where each bottleneck manifests in a different part of the network, both locations are correctly identified by the corresponding point of analysis.
Once again, by combining this information we can accurately detect the complex composite bottleneck.

Comparatively, as shown in Table \ref{table:4}, both centralized and distributed approaches behave similarly when the composite bottleneck affects the same part of the network. On the other hand, in the cases where each distributed point of analysis only needs to deal with one of the bottleneck components, it can effectively identify the complete composite bottleneck while the centralized analysis can detect part of the problem.

{\em \textbf{Takeaway:}
Consideration of composite bottlenecks confirms that the distributed approach naturally lends itself to more accurately identifying bottlenecks that affect both the RAN and CN.
However, it still faces difficulties classifying composite bottlenecks that create complex problems within the same domain.
Going forward, we plan to explore whether a finer grained distribution can perform better when presented with complex bottlenecks as well as consider composite bottlenecks consisting of 3 or more individual bottlenecks.
}

\begin{table}
\small
\begin{tabular}{ | C{1.8cm} | C{1.8cm} | C{1cm} | C{1cm} | C{1cm} | } 
\hline
\multirow{2}{*}{\textbf{Profile}} 
& \multirow{2}{*}{\textbf{Centralized}} 
& \multicolumn{3}{c|}{\textbf{Distributed}} \\ \cline{3-5}
& & \textbf{Final} & \textbf{RAN} & \textbf{CN} \\ 
\hline
18 & \cellcolor[HTML]{ffffbf}50\% & \cellcolor[HTML]{a1d76a}100\% & \cellcolor[HTML]{ffffbf}50\% & \cellcolor[HTML]{ffffbf}50\% \\
\hline
19 & \cellcolor[HTML]{ffffbf}50\% & \cellcolor[HTML]{ffffbf}50\% & \cellcolor[HTML]{ffffbf}50\% & \cellcolor[HTML]{fc8d59}0\% \\
\hline
20 & \cellcolor[HTML]{ffffbf}50\% & \cellcolor[HTML]{a1d76a}100\% & \cellcolor[HTML]{ffffbf}50\% & \cellcolor[HTML]{ffffbf}50\% \\
\hline
\end{tabular}
\vspace{1mm}
\caption{Performance of composite bottleneck identification for the centralized and distributed approaches, in percentage of the bottleneck components.}
\label{table:4}
\end{table}
\section{Discussion}
\label{sec:discussion}

In this section, we discuss our observations regarding the utility of our measurements for automating bottleneck identification and attribution. In addition, we consider the performance of our approaches regarding computational and communication overhead. Finally we consider the limitations of our work and practical implications.

\subsection{The right set of measurements}

In Table~\ref{table:5}, we take another look at the measurements, this time to examine which ones are picked out by the automated feature selection process for the different bottleneck identification approaches.

We find that 6 measurements stand out.
The CQI, which is one of two available measurements (the other being radio RTX) that can clearly identify interference. 
The MCS, which while not easy to understand through simple (e.g. visual) inspection as shown in Sec.~\ref{sec:analysis}, can provide indication of any of the bottlenecks (see Table~\ref{table:1}) and proves to be useful for automating identification. 
The TCP RTX, which can clearly identify packet-loss. 
Memory usage, which is one of three measurements (i.e. the other two are storage and CPU utilization) that can clearly identify a lack of computational resources. 
We believe that one of three resource oriented measurements was sufficient to point to infrastructure stress because all computational resources were stressed equally. 
We expect that for more complex stress, the other two measurements would play an equally important part in identifying the computational resource bottlenecks. 
The TX/RX measurements, which can clearly identify congestion. 
Finally, link delay, which can help identifying delay-related bottlenecks. 

In summary, measurements from both the network function and the infrastructure layers are utilized, with one well targeted measurement playing the key role in identifying each bottleneck. 
This might lead one to believe that reverting to a handful of selected measurements is a straightforward decision and even simple approaches like visual inspection could prove fruitful on such a restricted set.
However, the limitations discussed in section~\ref{sec:analysis_initial} are still valid. 
The value of the targeted measurements can be quickly obscured by composite bottlenecks and in addition parameters like the MCS, which was of significant importance to all approaches and bottlenecks, are extremely difficult to decode visually or by simple threshold-based approaches.
Finally, the system size, data volume and correlations between different VNFs which are not considered in this work will quickly dictate the need for smart approaches using ML techniques.

\begin{savenotes}
\begin{table}
\small
\begin{tabular}{ | m{1.5cm} | M{1cm} | M{1.45cm} | M{1.45cm} | M{1.45cm} | }
 \cline{2-5}
    \multicolumn{1}{c|}{}
 & \textbf{Initial \newline analysis} & \textbf{Centralized \newline 5-clusters} & \textbf{Centralized \newline 9-clusters} & \textbf{Distributed} \\
\hline
\rowcolor[HTML]{fc8d59}
Throughput & & & & \\
\hline
\rowcolor[HTML]{fc8d59}
RTT & & & & \\
\hline
\rowcolor[HTML]{ffffbf}
Radio TX & & & & \\
\hline
\rowcolor[HTML]{ffffbf}
Radio RTX & & & & \\
\hline
\rowcolor[HTML]{ffffbf}
CQI & & \checkmark & \checkmark & \checkmark \\
\hline
\rowcolor[HTML]{ffffbf}
MCS & \checkmark  & \checkmark  & \checkmark  & \checkmark \\
\hline
\rowcolor[HTML]{ffffbf}
MSD & & & \checkmark & \\
\hline
\rowcolor[HTML]{ffffbf}
TCP RTX & \checkmark & & \checkmark  & \checkmark \\
\hline
\rowcolor[HTML]{91bfdb}
CPU & & & & \checkmark \\
\hline
\rowcolor[HTML]{91bfdb}
Memory & \checkmark & \checkmark & \checkmark & \checkmark \\
\hline
\rowcolor[HTML]{91bfdb}
Storage & & & & \\
\hline
\rowcolor[HTML]{91bfdb}
TX/RX & \checkmark & \checkmark & \checkmark & \checkmark \\
\hline
\rowcolor[HTML]{91bfdb}
Link Delay  & \checkmark & \checkmark & \checkmark & \checkmark \\
\hline
\end{tabular}
\vspace{1mm}
\caption{Measurements selected by the feature selection process for each of the approaches presented in Sections~\ref{sec:analysis}~and~\ref{sec:bottlenecks}}
\label{table:5}
\end{table}
\end{savenotes}

\subsection{Overhead}
\label{sec:discussion_overhead}

Bottleneck identification systems need to balance accuracy and overhead in terms of compute resources and strain on networking infrastructure.
Here, we explore whether the proposed approaches can provide such balance. 
Specifically we look at the centralized and distributed approaches categorizing by bottleneck type and location.
Because the approaches are not yet implemented, we do not have measurements that can quantify the trade-off precisely.
We, therefore, make a number of assumptions in estimating overhead ``units``.  
We divide the overhead to two categories, data-transfer and processing.
These are roughly estimated as follows.
Each distinct measurement carries a base overhead of 1 unit for processing and 1 unit for data-transfer.
This is meant to capture the relative contribution of each category.
Remote measurements double the data-transfer overhead considering that the data needs to be shipped to the central analysis point.
The processing overhead re-occurs for each processing step (i.e. each time clustering is performed).
Recall that our goal here is not to precisely quantify the involved overhead, but rather describe it qualitatively and thus the assumptions above can be viewed as an attempt to capture the various components of overhead.

Based on the above assumptions, each measurement initially produces the same overhead. 
The difference between the two approaches stems from the way that data is shipped and processed.
The centralized approach produces higher data-transfer overhead since the remote measurements need to reach the central analysis point. 
On the other hand, the distributed approach avoid the data-transfer overhead by processing both in the RAN and CN analysis points, producing higher processing overhead per measurement.
An opportunity for improving the overhead of the distributed approach lies in intelligently invoking specific analysis points only when they are needed (e.g. first attempt to identify the bottleneck in the RAN, if it is identified with high confidence there is no need to invoke analysis in the CN).

\begin{figure}[t]
	\centering
	\includegraphics[width=1\columnwidth]{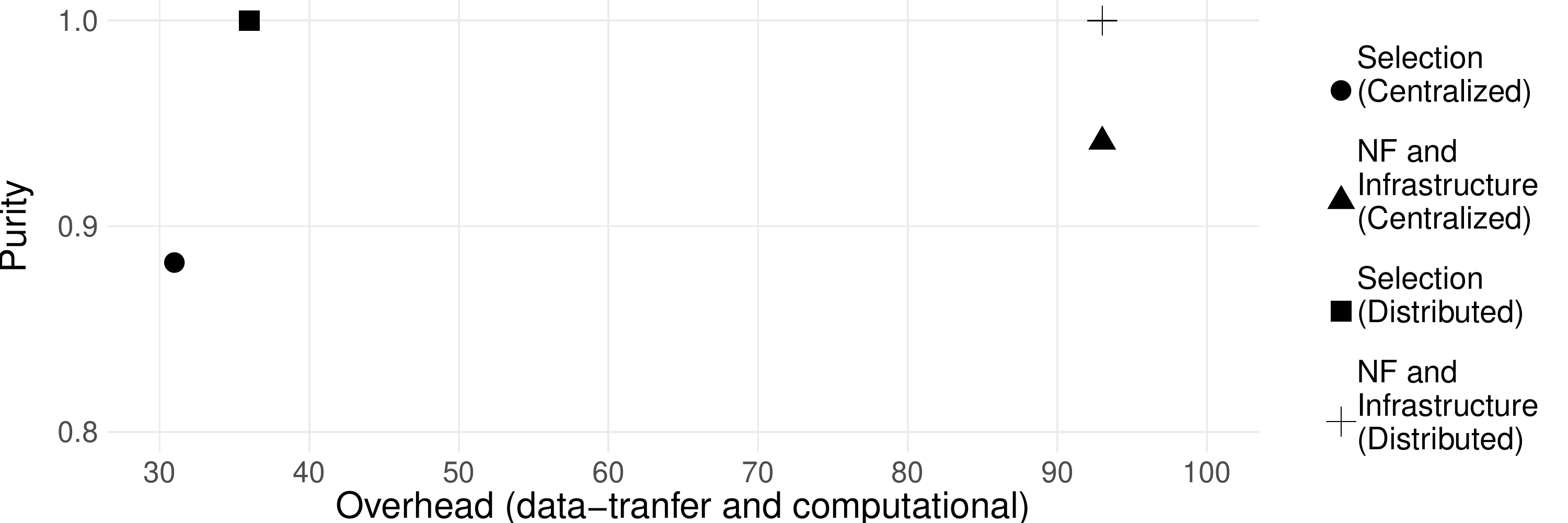}
	\vspace{-6mm}
	\caption{The purity obtained by the tested methodologies and best-performing feature-sets is contrasted to the combined data transfer and computational overhead that these bring.}
	\label{fig:overhead_purity}
\end{figure}

Using the simple assumption above, we estimate the overhead of the centralized and distributed approaches presented in Sec.~\ref{sec:bottlenecks}.
The obtained purity is shown in Fig.~\ref{fig:overhead_purity} as a function of the estimated overhead for the examined identification approaches and feature-sets.
There are two distinct groups based on overhead, owing to the feature-sets used. 
Feature selection, as expected, helps achieve a significant reduction in overhead, a $\approx$60\%-65\% reduction compared to the larger feature-set, regardless of the identification method.
Dissecting this overhead to its components, the picture is identical for both feature-sets. 
The centralized approach owes $\approx$33\% to data-transfer and $\approx$66\% to processing, while for the distributed approach the ratio is reversed with $\approx$66\% of the overhead being data-transfer and $\approx$33\% processing.
Another interesting observation is that the distributed approach consistently achieves perfect purity which should make the slight increase in overhead an acceptable cost.

Although, we cannot precisely estimate the involved communication and processing overhead, we can draw upon previous LTE measurements to get a sense of measurement data volumes.
Iyer et al.~\cite{iyer2017automating} collected RAN metrics, a set that closely matches what we have we collected, from 14,000 basestations in an operational setting along with RLC traffic.
They reported traffic volumes of 6 TeraBytes (TB) per hour and summary measurements data of 100s of TB per year.
These numbers hint that fully instrumenting an operational RAN (i.e. collecting both measurements and packet traces) can result on collecting and processing a few TBs of data every hour and exchanging summary information that can amount to 100s of GBs. 
This indicates a potentially non-trivial strain on compute and networking resources, which strengthens the case for intelligent measurement selection as well as distributed monitoring, anomaly detection and root cause analysis approaches.

\subsection{Limitations}
This work has certain limitations related to the scale of the testbed and the generalizability of the results.

At the time that this work was completed, the deployment of commercial 5G networks was still at its infancy and the availability of 5G data (e.g. from trial or operational networks), testbeds (e.g. 5GVINNI~\cite{5gvinni}, POWDER~\cite{powder}) and simulators (e.g. ns-3~\cite{ns3}) was non-existent or extremely limited.
With this in mind we set up our own testbed as described in Sec.~\ref{sec:methodology_testbed} that would be readily available for experimentation.

Despite the small scale and the use of 4G components, we believe this testbed to be entirely sufficient for the scope of this work which is to empirically highlight the key challenges of instrumenting and monitoring a cloudified mobile network architecture, such as the one envisioned by 5G, for the purposes of bottleneck identification. 
It successfully brings together the domains of data-center infrastructure, virtualization of network functions and the next generation mobile network to provide interesting insights and a solid foundation for future research.

\subsection{Lessons learned and practical implications}

This exercise has highlighted the intractability of the task of bottleneck identification and attribution in a cloudified mobile network architecture.

However, we found that distinct bottleneck signatures can be identified for a wide variety of performance degradations, provided that sufficient data from multiple perspectives can be collected and intelligently processed.
Unsupervised classification of bottlenecks by signature is possible and makes it easier to understand what the nature and location of the bottleneck is. 
However, it does not readily translate to a fully automated system.
While in this work we were in control of the bottleneck profiles, understanding the classification of bottlenecks encountered in the wild, in an operational network will require further intervention (e.g. labeling of anomalies).
The success of unsupervised learning in telling apart different bottlenecks is encouraging and suggests that a properly trained and tuned supervised classifier can potentially yield perfect results. 
Training such models in a dynamic network remains challenging and an interesting research direction.
Multi-layered methodologies can also be envisioned where fast anomaly detection~\cite{ahmed2016survey} is the first line of defence, while bottleneck localization and identification is deferred to secondary analysis.
We also find that a centralized monitoring system can face difficulties when encountering complex bottlenecks.
While methods for concurrently identifying multiple bottlenecks exist, the distributed architecture can naturally tackle the task from multiple points of view while maintaining a low overhead, giving it a notable edge.
However, designing an efficient distributed monitoring system mandates solving a number of key problems.
These include the distribution granularity and how to coordinate between its compartments.
The finer the granularity the better the system at attributing composite bottlenecks.
A finer granularity however, would complicate the task of coordinating between the different compartments in the system and may impact the timeliness of bottleneck identification.
We plan to explore these trade offs in our future work.
\section{related work}
\label{sec:related}

Based on our analysis and efforts in bottleneck identification there is strong evidence that holistic cross-layer monitoring solutions, exploiting a rich set of monitoring data are of great importance for identifying potential bottlenecks and providing service performance guarantees in the emerging cloudified mobile systems. However, considering the state of the art in the monitoring space, we can observe a disparity in the way that the problem is dealt with depending on the domain, both in terms of industrial solutions and the research literature.

In the domains of cloud computing and data centers, a plethora of practical monitoring systems already exist, either tailored for specific environments (e.g. Microsoft's Azure Monitor~\cite{azure-mon} for the Azure cloud) or targeting more generic infrastructure deployments, like Zabbix~\cite{zabbix} and Nagios~\cite{nagios}. Such systems allow the monitoring of KPIs like CPU/network utilization, memory/storage usage, network flow information etc, both in terms of the underlying physical infrastructure and of the deployed virtual network functions and support the triggering of events and alarms whenever certain important KPIs cross some threshold. In accordance to this, the focus of research works in this space is on solutions that can make the monitoring of the relevant resources as efficient and as fine grained as possible, either from the angle of cloud computing (e.g.~\cite{calero2015monpaas}) or from the angle of the data center and the efficient monitoring of network flows (e.g.~\cite{li2016flowradar,moshref2016trumpet}). 
However, in all these cases monitoring is treated generically, not taking into consideration domain specific monitoring information originating at the network function and service layers, which as shown in Sections \ref{sec:analysis} and \ref{sec:bottlenecks}, can improve the identification of the root cause of performance bottlenecks. 

In an analogous manner, monitoring solutions exist in the mobile domain, like Nokia's Wireless Network Guardian~\cite{nokia-wng} and Amdocs' Deep Network Analytics~\cite{amdocs}. 
In contrast to the cloud solutions, these focus on analyzing data from the mobile networking domain such as RAN  and CN related information like the ones in this work, to identify potential bottlenecks. In this case, the type of bottlenecks that are being considered are different in nature and revolve around service-related issues, like interference, sudden traffic surges, etc. This focus can also be seen in the research literature, with some examples being~\cite{iyer2015celliq,baranasuriya2015qprobe}, which propose ways of exploiting mobile network information to identify bottlenecks in real-time.

A number of new proprietary products are starting to appear that attempt to fuse data from both domains (e.g. Ericsson's Network Manager~\cite{ericcson-nm}). Similarly, both ETSI OSM~\cite{etsi2016open} and ONAP~\cite{onap-2018}, the most prominent solutions for the management and orchestration of cloudified mobile networks, provide subsystems that can enable multi-source monitoring.

There is a fair amount of previous work on bottleneck characterization and monitoring of traditional mobile networks (e.g.~\cite{rengaraju2012qoe,baranasuriya2015qprobe,iyer2017automating}) and cloud computing infrastructures (e.g.~\cite{calero2015monpaas,li2016flowradar,moshref2016trumpet,cao2015nfv,naik2016nfvperf,Kandula:2009:DDE:1592568.1592597}).
Equally, there is work on bottleneck characterization in virtualized mobile networks but focusing solely on the CN (e.g.~\cite{rajan2015understanding,prados2017modeling,raza2017rethinking}).

On the side of advanced monitoring intelligence, z-TORCH~\cite{sciancalepore2018z}, is an automated NFV orchestration and monitoring solution for generic virtual network functions that provides an adaptive monitoring mechanism in terms of the data collection frequency, and attempts to profile the behavior of VNFs. On a parallel path,~\cite{PadmanabhaIyerTradeoff}, examines the trade-off of latency and accuracy in the domain of mobile analytics, focusing on the RAN. 

To the best of our knowledge, there is no work that seeks to holistically and experimentally examine the range of performance bottlenecks that can impact service quality in a cloudified mobile network setting along with the measurement parameters that can help identify them.
While capturing multi-layer data is already possible, the exact type of data that should be captured at any point in time and their monitoring frequency is still unclear.
Moreover, as our analysis showed, the link between cause and effect of performance degradation becomes fuzzier due to the complexity of the network and the virtualization of mobile network functions, complicating the monitoring process further. Also, naive approaches to centrally collecting all monitoring data are unlikely to scale in operational networks~\cite{meng2011state, meng2012reliable, povedano2013dargos, onapmultisite}.

In order to resolve the aforementioned challenges, a more intelligent monitoring solution is required, that draws a balance in terms of scalability and accuracy of identifying bottlenecks that lead to QoS degradation. Exploiting insights like the ones obtained in Sections \ref{sec:analysis} and \ref{sec:bottlenecks} regarding the usefulness of the various features is a good step towards this direction.

An effective framework targeting performance assurance in a cloudified mobile network context, should build on such ideas and expand their scope to capture all of the aspects solidified by the 5G system architecture, including both the domains of cloud computing and mobile networks and the idiosyncrasies of the individual network functions that are involved.

\section{Conclusions}
\label{sec:conclusions}

In this work, we have presented an experimental study with the goal of characterizing cloudified mobile network performance bottlenecks and explored the challenges in identifying them. 
To achieve this, we employed a prototype testbed that allows the creation of end-to-end mobile network slices through the NFV paradigm. 
Our experiments demonstrated the complexity involved in identifying such bottlenecks even for simple scenarios and revealed how obtaining monitoring data from various layers of the cloudified ecosystem can improve this identification process. 
Based on the insights gained from this exercise and considering the currently available monitoring solutions, it is clear that a novel and more intelligent monitoring framework is required for the assurance of 5G service performance guarantees, taking into consideration both the accuracy of identifying bottlenecks and the overhead of monitoring.
This is exemplified by the consideration of complex scenarios featuring composite bottlenecks.
Our evaluations show that a decentralized bottleneck identification approach offers high accuracy while keeping overhead reasonable.    
Designing a monitoring system that meets the aforementioned requirements and expanding our current experimentation framework to capture more complex bottlenecks with more types of monitoring data are two important dimensions to consider in our future work.

\bibliographystyle{IEEEtran}
\bibliography{IEEEabrv,main}

\end{document}